\documentclass[11pt,a4paper]{article}
\usepackage{jheppub}

\usepackage{slashed}
\usepackage{epsfig}

\newcommand{\vect}[1]{\boldsymbol{#1}}
\def\vb{\begingroup\obeyspaces\u}
\def\u#1{\tt#1\endgroup}

\title{Single Perturbative Splitting Diagrams in Double Parton Scattering} 
\author{Jonathan R. Gaunt$^{a,b}$} 
\affiliation{$^a$Cavendish Laboratory, University of Cambridge, J.J. Thomson Avenue, 
\\Cambridge
CB3 0HE, U.K. \\
$^b$Deutsches Elektronen-Synchroton DESY, 22603 Hamburg, Germany} 

\emailAdd{gaunt@hep.phy.cam.ac.uk} 
\abstract{We present a detailed study of a specific class of graph that can potentially contribute to the proton-proton
double parton scattering (DPS) cross section. These are the `2v1' or `single perturbative splitting' graphs, in which two 
`nonperturbatively generated' ladders interact with two ladders that have been generated via a perturbative $1 \to 2$ 
branching process. Using a detailed calculation, we confirm the result written down originally by Ryskin and Snigirev -- namely, 
that the 2v1 graphs in which the two nonperturbatively generated 
ladders do not interact with one another do contribute to the leading order proton-proton DPS cross section, albeit with a different geometrical
prefactor to the one that applies to the `2v2'/`zero perturbative splitting' graphs. We then show that 2v1 graphs
in which the `nonperturbatively generated' ladders exchange partons with one another also contribute to the 
leading order proton-proton DPS cross section, provided that this `crosstalk' occurs at a lower scale than the $1 \to 2$ branching 
on the other side of the graph. Due to the preference in the 2v1 graphs for the $x$ value at which the branching occurs, and
crosstalk ceases, to be very much larger than the $x$ values at the hard scale, the effect of crosstalk interactions is likely to
be a decrease in the 2v1 cross section except at exceedingly small $x$ values ($\lesssim 10^{-6}$). At moderate $x$ values
 $\simeq 10^{-3} - 10^{-2}$, the $x$ value at the splitting is in the region $\simeq 10^{-1}$ where PDFs do not change much with scale, and 
the effect of crosstalk interactions is likely to be small.  We give an explicit
formula for the contribution from the 2v1 graphs to the DPS cross section, and combine this with a suggestion that
we made in a previous publication, that the `double perturbative splitting'/`1v1' graphs should be completely removed
from the DPS cross section, to obtain a formula for the DPS cross section. It is pointed out that there are two
potentially concerning features in this equation, that might indicate that our prescription for handling the 
1v1 graphs is not quite correct.}


\dedicated{Cavendish--HEP--12/11}

\begin{document}

\maketitle

\section{Introduction} \label{sec:Xsecintro}

We define double parton scattering (DPS) as the process in which two pairs of partons participate in hard interactions in a 
single proton-proton (p-p) collision. This process is formally suppressed relative to the usual single parton scattering (SPS)
mechanism by $\Lambda^2/Q^2$, where $Q^2$ is the scale of the hard scattering(s) involved, and $\Lambda^2$ is the scale of
nonperturbative physics. However, DPS can contribute important backgrounds to SPS processes that are suppressed by small or 
multiple coupling constants, such as Higgs or new physics signals \cite{DelFabbro:1999tf, Hussein:2006xr, Hussein:2007gj,
Bandurin:2010gn}. It is also an interesting process to study in its own right, as it reveals novel information concerning
correlations between partons in the proton. For any final state $AB$ which can potentially have been produced via two
independent scatterings yielding $A$ and $B$, there is a region of final state phase space in which the DPS production
mechanism is competitive with the SPS one -- namely, the region in which the transverse momenta of $A$ and $B$ are small
\cite{Diehl:2011tt, Diehl:2011yj}. This fact offers some scope to study DPS in detail -- indeed it has been used in past
experimental extractions of DPS \cite{Akesson:1986iv, Alitti:1991rd, Abe:1993rv, Abe:1997xk, Abazov:2009gc, 
ATLAS-CONF-2011-160}. Finally, the rate of DPS relative to SPS for a given final state $AB$ increases with collider 
energy, as lower $x$ values are probed where the population of partons is larger. This means that DPS will be more 
important at the LHC than at any previous collider. For these reasons, there has been a considerable increase in 
interest in the phenomenon of DPS in recent years from both the experimental and theoretical communities, and 
four international workshops have been organised on the theme \cite{Bartalini:2010su, MPI10page, MPILHC10page, 
MPI11page, Bartalini:2011jp}.

If one assumes that the two hard processes $A$ and $B$ may be factorised, then the total cross section for the
process $pp \to AB + X$ via double parton scattering should be of the following form:
\begin{align} \label{DPSXsecgen}
\sigma^{D}_{(A,B)} \propto& \sum_{i,j,k,l}\int \dfrac{d^2\vect{r}}{(2\pi)^2}\prod_{a=1}^{4}dx_a \Gamma_{ij}(x_1,x_2,\vect{r};Q_A^2,Q_B^2) \Gamma_{kl}(x_3,x_4,-\vect{r};Q_A^2,Q_B^2)
\\ \nonumber
\times& \hat{\sigma}_{ik \to A}(\hat{s} = x_1x_3s) \hat{\sigma}_{jl \to B}(\hat{s} = x_2x_4s)
\\ \nonumber
\propto& \sum_{i,j,k,l}\int d^2\vect{b}\prod_{a=1}^{4}dx_a \Gamma_{ij}(x_1,x_2,\vect{b};Q_A^2,Q_B^2) \Gamma_{kl}(x_3,x_4,\vect{b};Q_A^2,Q_B^2)
\\ \nonumber
\times& \hat{\sigma}_{ik \to A}(\hat{s} = x_1x_3s) \hat{\sigma}_{jl \to B}(\hat{s} = x_2x_4s)
\end{align}
The $\hat{\sigma}$ symbols represent parton-level cross sections. $\Gamma_{ij}(x_1,x_2,\vect{b};Q_A^2,Q_B^2)$
is the impact-parameter space two-parton GPD ($\vect{b}$-space 2pGPD), whilst $\Gamma_{ij}(x_1,x_2,\vect{r};Q_A^2,Q_B^2)$
is the transverse momentum space 2pGPD ($\vect{r}$-space 2pGPD). $\Gamma_{ij}(x_1,x_2,\vect{b};Q_A^2,Q_B^2)$ has a 
probability interpretation as the probability to find a pair of quarks in the proton with flavours $ij$, momentum 
fractions $x_1x_2$, and separated by impact parameter $\vect{b}$, at scales $Q_A$ and $Q_B$ respectively 
\cite{Blok:2010ge, Diehl:2011tt, Diehl:2011yj}. The $\vect{r}$-space 2pGPD is the Fourier transform of 
this with respect to $\vect{b}$, and has no probability interpretation. $\vect{r}$ is related to the 
transverse momentum imbalance of one of the partons emerging from the proton between amplitude and 
conjugate (for more detail, see Section 2 of \cite{Diehl:2011tt}).

Since the experimental extraction of DPS relies on the fact that the DPS cross section differential
in the transverse momenta of $A$ and $B$, $\vect{q_A}$ and $\vect{q_B}$, is strongly peaked at small 
$\vect{q_A}$ and $\vect{q_B}$, it is perhaps the DPS cross section differential in $\vect{q_A}$ and 
$\vect{q_B}$ rather than the total cross section that is more relevant for making experimentally 
testable predictions \cite{Blok:2010ge, Blok:2011bu, Diehl:2011tt, Diehl:2011yj}. In the region of $\vect{q}_A, 
\vect{q}_B$ of interest (i.e. $\vect{q}_A^2, \vect{q}_B^2 \ll Q_A^2, Q_B^2$) this quantity is described 
in terms of transverse momentum dependent 2pGPDs (TMD 2pGPDs), rather than the collinear 2pGPDs 
appearing in \eqref{DPSXsecgen}. On the other hand, it is expected that for $\Lambda^2 \ll \vect{q}_A^2, 
\vect{q}_B^2 \ll Q_A^2, Q_B^2$ the TMD 2pGPD should be expressible in terms of the collinear 2pGPD 
and a perturbatively calculable piece \cite{Diehl:2011tt, Diehl:2011yj}. In that case there is a 
`collinear part' of the differential cross section whose structure closely resembles the total cross 
section formula \eqref{DPSXsecgen}. Knowledge of how the total DPS cross section is to be treated 
should be helpful in establishing the correct way to treat this collinear part. It is with this 
ultimate purpose in mind that we continue to discuss only the total cross section for DPS in the 
remainder of the paper.

\begin{figure}
 \centering
 \includegraphics[scale=0.4]{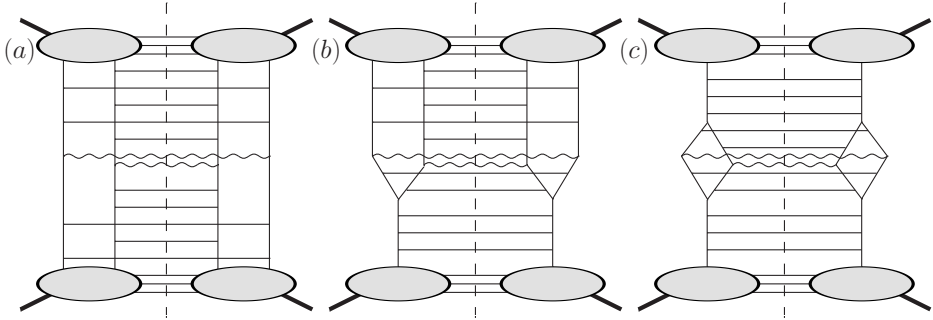}
 \caption{Some types of graph that can potentially contribute to the DPS cross section.}
 \label{fig:2v21v2}
\end{figure}

There are several classes of graph that can potentially contribute to the leading order (LO) p-p DPS 
cross section (we restrict our attention to leading order, or leading logarithmic accuracy, in this
paper). These are sketched in figure \ref{fig:2v21v2}. Note that the partons emerging from the grey 
proton blobs in the figure are nonperturbatively generated partons -- i.e. ones existing at a low 
scale $\sim \Lambda_{QCD}$ -- and that we've taken all the hard processes in the figure to be the
production of an electroweak gauge boson with positive invariant mass (denoted by a wiggly line).
We'll refer to the different classes of graph (a), (b), and (c) as 2v2, 2v1 and 1v1 graphs 
respectively, for obvious reasons.

In the paper \cite{Gaunt:2011xd}, we carefully examined graphs of the 1v1 type. We found that the 
treatment of these graphs by a long-established framework for calculating the p-p DPS cross section
\cite{Shelest:1982dg, Zinovev:1982be, Snigirev:2003cq} was unsatisfactory, and suggested that no part
of these graphs should be included as part of the leading order p-p DPS cross section.
Since then, this suggestion has also been made in a number of other papers \cite{Blok:2011bu, Manohar:2012pe}.

In light of this discovery, a careful re-analysis of other classes of graph that can potentially contribute
to the LO DPS cross section would seem appropriate. In this paper we will pay particular attention to the 2v1 graphs
in which there is only a single perturbative splitting, such as that drawn in figure \ref{fig:2v21v2}(b) 
(we'll also discuss to a certain extent 2v2 graphs such as \ref{fig:2v21v2}(a) in which there are no perturbative
splittings, although it should be reasonably clear that these should be included in the LO DPS cross section).

\begin{figure}
 \centering
 \includegraphics[scale=0.7]{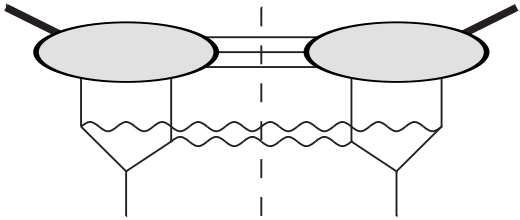}
 \caption{The simplest structure possible for the 2v1 graph.}
 \label{fig:1v2simplest}
\end{figure}

In section \ref{sec:2v1calc}, we will begin to address the issue of whether contributions from the 2v1 graphs
should be included in the LO DPS cross section, and what form these contributions should take. We'll do this 
using a similar strategy as we employed for the 1v1 graphs in the paper \cite{Gaunt:2011xd}. That is, we'll take a 2v1 
graph with the simplest possible structure (i.e. the structure of figure \ref{fig:1v2simplest}) and see 
whether there is a `natural' part of the cross section expression for it that is proportional to $1/R_p^2$ ($R_p =$ proton radius), 
and also contains a large logarithm associated with the $1 \to 2$ splitting. The large logarithm should be 
associated with transverse momenta of the partons emerging from the $1 \to 2$ splitting being $\ll Q^2$ 
(where we take $Q_A^2 = Q_B^2 \equiv Q^2$ for simplicity). If there is such a structure in the 2v1 graph, 
then this part of this graph should be included in the LO DPS cross section. Furthermore, if there is a 
$\log(Q^2/\Lambda^2)/R_p^2$ structure in the simplest 2v1 diagram, then we expect there to be a 
$\log(Q^2/\Lambda^2)^n/R_p^2$ piece in the more general 2v1 diagram of figure \ref{fig:2v21v2}(b) that 
should also be included in the LO DPS cross section. This will be associated with the branchings in the 
diagram being strongly ordered in transverse momentum. From the structure of the contribution to the
LO DPS cross section coming from the simplest 2v1 diagram, we'll be able to write down a resummed expression 
for the contribution to the LO DPS cross section coming from 2v1 diagrams with the structure of figure
\ref{fig:2v21v2}(b).

\begin{figure}
\centering
\includegraphics[scale=0.65]{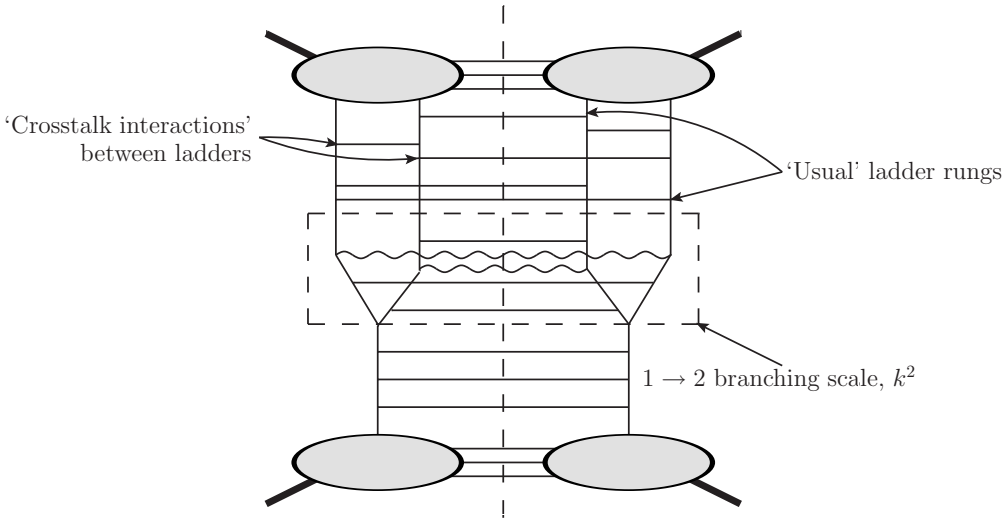}
\caption{Generic 2v1 diagram including `crosstalk' that we argue contributes to the 2v1 DPS cross section at the 
leading logarithmic level.}
\label{fig:genericcrosstalk}
\end{figure}

The results that we obtain in section \ref{sec:2v1calc} have in fact already been written down in the papers 
\cite{Ryskin:2011kk, Blok:2011bu, Manohar:2012jr, Manohar:2012pe}, and one can view the content of that section as a more detailed re-derivation 
of some of the results in those papers. In section \ref{sec:crosstalk}, we will however establish a further result
with regard to the contribution of 2v1 graphs to the LO DPS cross section. We will discover that the final formula
that we obtained in section \ref{sec:2v1calc} is incomplete, and that there are further diagrams of the 2v1 type
that contribute to the LO DPS cross section. These diagrams involve non-diagonal crosstalk interactions between the two 
nonperturbatively generated parton ladders, at scales lower than the perturbative $1 \to 2$ ladder branching
on the other side -- an example diagram of this type is sketched in figure \ref{fig:genericcrosstalk} (note that
there are no such crosstalk interactions in figure \ref{fig:2v21v2}(b)). This result is again established by
analysis of the simplest Feynman graphs of the appropriate type. In section \ref{sec:crosstalk} we'll also discuss 
in detail the issue of colour in relation to the crosstalk interactions, and make some comments with regard to the
potential numerical impact of the crosstalk interactions on the 2v1 DPS cross section.

In section \ref{sec:XsecResults}, we combine the results from sections \ref{sec:2v1calc} and 
\ref{sec:crosstalk} with our suggestion from \cite{Gaunt:2011xd} with regards to the 1v1 graphs,
to give a suggested expression for the LO cross section for DPS. We do however point out two
potentially concerning features in this equation that may indicate that the suggestion of 
\cite{Gaunt:2011xd} to completely remove the 1v1 graphs from the DPS cross section is not quite
the correct prescription.

\section{`Two versus One' Contributions to the DPS Cross Section} \label{sec:2v1calc}

\begin{figure}
\centering
\includegraphics[scale=0.7]{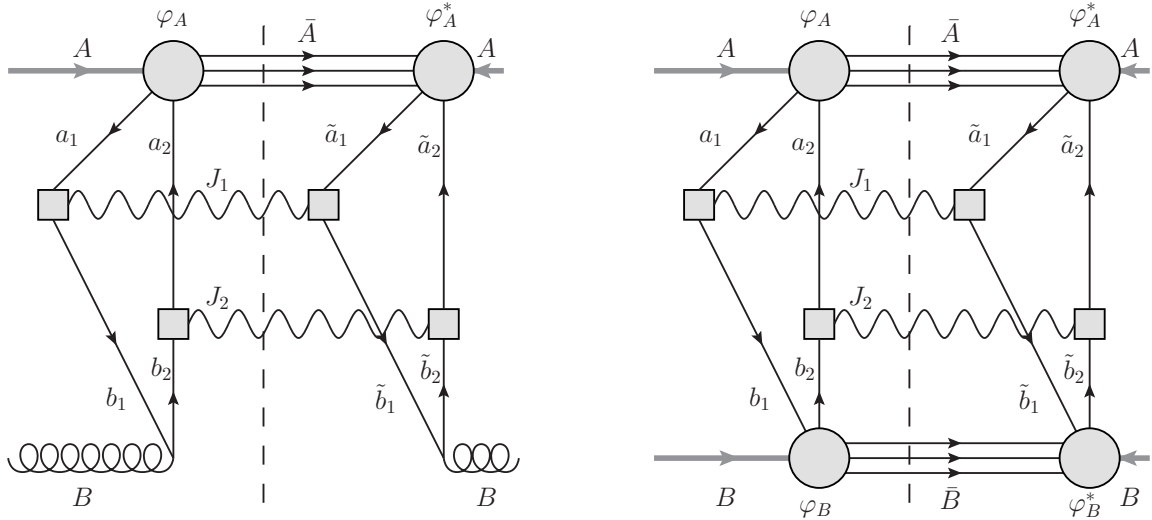}
\caption{(a) An example of a `2v1' DPS-type scattering diagram. (b)
An example of a `2v2' DPS-type scattering diagram. The thick grey lines
are protons, whilst the grey circles are proton vertices. The labels on the
lines correspond to the four momenta of those lines.}
\label{fig:DPSdiagrams}
\end{figure}

In this section, we show explicitly that for a 2v1 diagram with the structure of figure \ref{fig:1v2simplest},
there is a part of the cross section expression that contains a DGLAP-type large logarithm and a factor of
order $1/R_p^2$, which should be considered as part of the LO DPS cross section. We present details of the calculation
only for the particular flavour-diagonal contribution to the $gp \to gq\bar{q}+X \to \gamma^*\gamma^*+X$ process presented
in figure \ref{fig:DPSdiagrams}(a), where the two off-shell photons both have a positive invariant mass. However, the 
general method outlined below can be applied to any diagram of the appropriate structure, and will always give a large 
logarithm provided that the corresponding process is allowed in the collinear limit (apart from issues of $J_z$ 
nonconservation at the splitting vertex).

In the calculation of the cross section for figure \ref{fig:DPSdiagrams}(a), we will have to include a wavefunction
factor or hadronic amplitude $\varphi$ to find two nonperturbatively generated partons in the 
proton, at the amplitude level in the calculation. It would be inappropriate to try and calculate a 2v1 cross 
section in a naive `fully parton-level' way omitting the proton at the top of the diagram because then one would have three 
particles in the initial state (whereas the standard framework for calculating a cross section requires two particles
in the initial state). Furthermore, by deleting the proton at the top of the diagram one would then be neglecting the important
fact that the two partons on this side are tied together in the same proton (as was pointed out in \cite{Blok:2011bu}).
The use of proton wavefunctions or hadronic amplitudes in the calculation of DPS-type graphs was discussed long ago in 
\cite{Paver:1982yp}, and has been discussed more recently in \cite{Diehl:2011yj, Blok:2011bu}. We utilise the approach
and notation of \cite{Paver:1982yp} in our work. That is, we assign a wavefunction factor $\varphi$ to the $p \to q\bar{q}X$
vertex that is assumed to be strongly damped for values of the parton transverse momentum and virtuality larger than the hadronic
scale $\Lambda \sim 1/R_P$ . In our case the factor $\varphi$ is a matrix in spinor space, and also carries a label $\chi$ that
describes the spins of all of the particles in $X$.

In the following we will take a number of steps to simplify the calculation as much as possible. First, we will largely 
ignore considerations of colour, and will suppress colour indices, factors and sums where they appear, in order to avoid
the proliferation of too many indices. Second, we will take take the four-momenta squared of the two off-shell photons 
to be the same, and refer to this common four-momentum squared as $Q^2$. Finally, we will take the colliding protons to
be unpolarised, as is the case for the colliding protons at the LHC.

As in \cite{Gaunt:2011xd}, we apply a Sudakov decomposition to all four-vectors used -- that is, we write an arbitrary
four-vector $V$ in terms of two lightlike vectors $p\equiv\tfrac{1}{\sqrt{2}}(1,0,0,1)$ and $n\equiv\tfrac{1}
{\sqrt{2}}(1,0,0,-1)$ and a transverse part denoted as $\vect{V}$:
\begin{equation}
V = V^+p+V^-n+\vect{V} 
\end{equation}

Rather than proceeding to calculate the cross section contribution from figure \ref{fig:DPSdiagrams}(a) directly, we instead begin by calculating
the cross section contribution $\sigma_{2v2}$ associated with the Feynman diagram in figure \ref{fig:DPSdiagrams}(b). In this diagram, 
two nonperturbatively generated quark-antiquark
pairs produced by colliding protons interact via two separate $q\bar{q} \to \gamma^*$ hard processes. We should be able to express the
cross section of this process in terms of `nonperturbatively generated parton pair' $\vect{r}$-space 2pGPDs $\Gamma(x_1,x_2;\vect{\Delta})$ 
and hard subprocess cross sections $\hat{\sigma}$ as follows:
\begin{align} \label{2v2XS}
\sigma_{2v2}(s) = & \int dx_1dx_2dy_1dy_2 \hat{\sigma}_{q\bar{q} \to \gamma^*}(\hat{s} = x_1y_1s)
\hat{\sigma}_{q\bar{q} \to \gamma^*}(\hat{s} = x_2y_2s)
\\ \nonumber
&\times  \int \dfrac{d^2\vect{\Delta}}{(2\pi)^2} \Gamma_p(x_1,x_2;\vect{\Delta}) \Gamma_p(y_1,y_2;\vect{-\Delta})
\end{align}

Helicity labels are omitted in the above schematic expression, but they will be included in the full calculation below. By using
the fact that the expression for the cross section must end up in this form, we can establish the connection between the vertex
factor  $\varphi$ and the `nonperturbatively generated parton pair' $\vect{r}$-space 2pGPD $\Gamma$. We shall need to make use 
of this relationship when we come to study figure \ref{fig:DPSdiagrams}(a).

Note that a calculation of $\sigma_{2v2}$ has already been performed by Paver and Treleani in \cite{Paver:1982yp}
for the case of spinless partons, and by Mekhfi \cite{Mekhfi:1983az} and Diehl, Ostermeier and Schafer \cite{Diehl:2011tt, Diehl:2011yj}
for the case of partons with spin. We follow closely the approach of Paver and Treleani, and our calculation of $\sigma_{2v2}$ can be 
considered as a brief review of the method in \cite{Paver:1982yp}.

We will neglect the proton mass with respect to the total centre of mass energy $\sqrt{s}$ and work in a frame in which
$A$ is proportional to $p$, whilst $B$ is proportional to $n$, $A = A^+p, B=B^-n$. One can directly write down the
following expression for the cross section contribution from figure \ref{fig:DPSdiagrams}(b), $\sigma_{2v2}(s)$:
\begin{align} \nonumber
\sigma_{2v2}(s) =& \dfrac{1}{2(2\pi)^{10}s} \sum_{\chi\gamma} \int d^4\bar{A} d^4\bar{B}  d^4J_1d^4J_2 \delta^{(4)}(\bar{A}+\bar{B}+J_1+J_2-A-B) \delta(J_1^2-Q^2)
\\ \label{2v2XS1}
&\times \delta(J_2^2-Q^2) \mathcal{M}^{\chi\gamma\mu_1\mu_2}(A,B;\bar{A},\bar{B},J_1,J_2) \mathcal{M}^{\chi\gamma\mu_1\mu_2}(A,B;\bar{A},\bar{B},J_1,J_2)^*
\end{align}
where:
\begin{align} \label{2v2ME}
\mathcal{M}^{\chi\gamma\mu_1\mu_2}&(A,B;\bar{A},\bar{B},J_1,J_2)
\\ \nonumber
\equiv& \int \dfrac{d^4a_1}{(2\pi)^4} \dfrac{Tr\left[T^{\mu_1}(J_1)\slashed{a}_1\varphi_p^\chi(a_1,a_2,\bar{A})\slashed{a}_2
T^{\mu_2}(J_2) \slashed{b}_2 \varphi_p^\gamma(b_2,b_1,\bar{B})\slashed{b}_1\right]}{D(a_1)D(a_2)D(b_1)D(b_2)},
\end{align}
\begin{align}
D(a) \equiv& a^2 + i\epsilon, \qquad
T^{\mu_1}(J_1) \equiv ieQ_q\slashed{\varepsilon}^{*}_{\mu_1}(J_1)
\end{align}
\begin{align}
a_2 \equiv A -\bar{A} - a_1 \qquad b_1 \equiv J_1 - a_1 \qquad b_2 \equiv B - \bar{B} + a_1 - J_1
\end{align}

The vertex factors $\varphi$ ensure that the quark and antiquark lines with momenta $a_i$ and $b_i$ have small virtuality.
Given that this is the case, we can rewrite the slashed vectors in \eqref{2v2ME} as sums over outer products of particle
or antiparticle spinors (as appropriate), using the completeness relations. Then we have:
\begin{align} \label{2v2ME2}
\mathcal{M}^{\chi\gamma\mu_1\mu_2}&(A,B;\bar{A},\bar{B},J_1,J_2)
\\ \nonumber
\simeq & \int \dfrac{d^4 a_1}{(2\pi)^4} \sum_{s_it_i}\mathcal{M}^{s_1t_1;\mu_1}_{q\bar{q} \to \gamma^*}(a_1b_1 \to J_1)
\mathcal{M}^{s_2t_2;\mu_2}_{\bar{q}q \to \gamma^*}(a_2b_2 \to J_2)
\\ \nonumber
& \times \left[\dfrac{\bar{u}^{s_1}(a_1) \varphi^\chi_p(a_1,a_2,\bar{A}) v^{s_2}(a_2)}{D(a_1)D(a_2)}\right]
\left[\dfrac{\bar{u}^{t_2}(b_2) \varphi^\gamma_p(b_2,b_1,\bar{B}) v^{t_1}(b_1)}{D(b_1)D(b_2)}\right]
\end{align}

The $s_i$ and $t_i$ are quark or antiquark helicity labels, and the $\mathcal{M}_{q\bar{q} \to \gamma^*}$ factors
are `hard' $q\bar{q} \to \gamma^*$ matrix elements. The hard matrix elements should be evaluated with initial
state partons having small (i.e. hadron scale) transverse momenta and off-shellness -- however, we
make the approximation in the matrix elements that the initial-state partons are on-shell and collinear,
which only corresponds to a small relative error $\mathcal{O}(\Lambda^2/Q^2) \ll 1$.

Consider now the integrations over the longitudinal parts of $a_1$ -- i.e. $a_1^+$ and $a_1^-$. It is not
hard to show that the integration over $a_1^-$ is restricted to values of order $\Lambda^2/Q$ by the
vertex factor $\varphi(a_1,a_2,\bar{A})$, whilst $\varphi(b_1,b_2,\bar{B})$, $D(b_i)$, and the 
$\mathcal{M}_{q\bar{q} \to \gamma^*}$ are practically constant in this range (and approximately
equal to their values with $a_1^-$ set to zero). Similarly, the integration over $a_1^+$ is restricted 
to values differing from $J_1^+$ by $\sim\Lambda^2/Q$ by the vertex factor $\varphi(b_1,b_2,\bar{B})$, 
with $\varphi(a_1,a_2,\bar{A})$, $D(a_i)$ and the $\mathcal{M}_{q\bar{q} \to \gamma^*}$ being approximately 
constant and equal to their values at $a_1^+=J_1^+$ in this range. This allows us to write:
\begin{align} \label{2v2ME3}
\mathcal{M}^{\chi\gamma\mu_1\mu_2}&(A,B;\bar{A},\bar{B},J_1,J_2)
\simeq  \sum_{s_it_i}\mathcal{M}^{s_1t_1;\mu_1}_{q\bar{q} \to \gamma^*}(J_1^+p,J_1^-n \to J_1)
\\ \nonumber
& \times \mathcal{M}^{s_2t_2;\mu_2}_{\bar{q}q \to \gamma^*}(J_2^+p,J_2^-n \to J_2) \int \dfrac{d^2 \vect{a_1}}{(2\pi)^2} \left[\int \dfrac{da^-_1}{2\pi}\dfrac{\bar{u}^{s_1}(a_1) \varphi^\chi_p(a_1,a_2,\bar{A}) v^{s_2}(a_2)}{D(a_1)D(a_2)}\right]_{a_1^+=J_1^+}
\\ \nonumber
& \times \left[\int \dfrac{db^+_1}{2\pi}\dfrac{\bar{u}^{t_2}(b_2) \varphi^\gamma_p(b_2,b_1,\bar{B}) v^{t_1}(b_1)}{D(b_1)D(b_2)}\right]_{b_1^- = J_1^-}
\end{align}

Define:
\begin{align} \label{wavdef}
\psi_{p;q\bar{q}}^{s_1s_2\chi}(a_1^+,a_2^+,\vect{a_1},\vect{a_2},\bar{A}^-) \equiv -\int \dfrac{da^-_1}{2\pi}\dfrac{\bar{u}^{s_1}(a_1) \varphi^\chi_p(a_1,a_2,\bar{A}) v^{s_2}(a_2)}{D(a_1)D(a_2)}
\end{align}

Then we can write $\mathcal{M}^{\chi\gamma\mu_1\mu_2}(A,B;\bar{A},\bar{B},J_1,J_2)$ in a more compact form:
\begin{align} \label{2v2ME4}
\mathcal{M}^{\chi\gamma\mu_1\mu_2}&(A,B;\bar{A},\bar{B},J_1,J_2)
\\ \nonumber
\simeq & \sum_{s_it_i} \mathcal{M}^{s_1t_1;\mu_1}_{q\bar{q} \to \gamma^*}(J_1^+p,J_1^-n \to J_1)
\mathcal{M}^{s_2t_2;\mu_2}_{\bar{q}q \to \gamma^*}(J_2^+p,J_2^-n \to J_2)
\\ \nonumber
& \times \int \dfrac{d^2 \vect{a_1}}{(2\pi)^2} \psi_{p;q\bar{q}}^{s_1s_2\chi}
(J_1^+,J_2^+,\vect{a_1},\vect{a_2},\bar{A}^-) \psi_{p;q\bar{q}}^{t_2t_1\gamma}(J_2^-,J_1^-,\vect{b_2},\vect{b_1},\bar{B}^+)
\end{align}

We now insert \eqref{2v2ME4} into \eqref{2v2XS1}, and make use of the following relation in the resulting expression:
\begin{align} \label{hardXS}
\mathcal{M}^{s_1t_1;\mu_1}_{q\bar{q} \to \gamma^*}(J_1^+p,J_1^-n \to J_1)&
\mathcal{M}^{*\tilde{s}_1\tilde{t}_1;\mu_1}_{q\bar{q} \to \gamma^*}(J_1^+p,J_1^-n \to J_1)(2\pi)\delta(J_1^2-Q^2)
\\ \nonumber
&= \hat{\sigma}^{s_1,t_1;\tilde{s}_1,\tilde{t}_1;\mu_1}_{q\bar{q} \to \gamma*}(\hat{s} = 2J_1^+J_1^-) 4J_1^+J_1^-
\end{align}

$\hat{\sigma}^{s_1,t_1;\tilde{s}_1,\tilde{t}_1;\mu_1}_{q\bar{q} \to \gamma*}$ is the $q\bar{q} \to \gamma^*$ `cross section' with $q,\bar{q},\gamma^*$
helicities $s_1, t_1, \mu_1$ in the matrix element, and $\tilde{s}_1,\tilde{t}_1,\mu_1$ in the conjugate matrix element (note that if $s_1 \neq \tilde{s}_1$
and/or $t_1 \neq \tilde{t}_1$ this is not a cross section in the strict sense). $\hat{\sigma}^{s_1,t_1;\tilde{s}_1,\tilde{t}_1;\mu_1}_{q\bar{q} \to \gamma*}$
is related to the spin-averaged ${q\bar{q} \to \gamma^*}$ cross section $\hat{\sigma}_{q\bar{q} \to \gamma^*}$ by:
\begin{align}
\hat{\sigma}^{s_1,t_1;\tilde{s}_1,\tilde{t}_1;\mu_1}_{q\bar{q} \to \gamma*} = 2\hat{\sigma}_{q\bar{q} \to \gamma*}\delta_{s_1,-t_1} \delta_{\tilde{s}_1,-\tilde{t}_1} 
\end{align}

The result of inserting \eqref{2v2ME4} into \eqref{2v2XS1} is:
\begin{align} \label{2v2XS2}
\sigma_{2v2}(s)& = \dfrac{1}{2(2\pi)^{12}s} \sum_{s_it_i\tilde{s}_i,\tilde{t}_i\chi\gamma} \int d^4\bar{A} d^4\bar{B} d^4J_1 \hat{\sigma}^{s_1,t_1;\tilde{s}_1,\tilde{t}_1;\mu_1}_{q\bar{q} \to \gamma*}(\hat{s} = 2J_1^+J_1^-) 4J_1^+J_1^-
\\ \nonumber
&\times  \hat{\sigma}^{s_2,t_2;\tilde{s}_2,\tilde{t}_2;\mu_2}_{q\bar{q} \to \gamma*}(\hat{s} = 2J_2^+J_2^-) 4J_2^+J_2^-   \int \dfrac{d^2\vect{a_1}}{(2\pi)^2} \dfrac{d^2\vect{\tilde{a}_1}}{(2\pi)^2}
 \psi_{p;q\bar{q}}^{s_1s_2\chi}(J_1^+,J_2^+,\vect{a_1},\vect{a_2},\bar{A}^-)
\\ \nonumber
&\times \psi_{p;q\bar{q}}^{t_2t_1\gamma}(J_2^-,J_1^-,\vect{b_2},\vect{b_1},\bar{B}^+) \psi_{p;q\bar{q}}^{*\tilde{s}_1\tilde{s}_2\chi}(J_1^{+},J_2^{+},\vect{\tilde{a}_1},\vect{\tilde{a}_2},\bar{A}^-) \psi_{p;q\bar{q}}^{*\tilde{t}_2\tilde{t}_1\gamma}(J_2^{-},J_1^{-},\vect{\tilde{b}_2},\vect{\tilde{b}_1},\bar{B}^+)
\displaybreak[3] \\ \nonumber
=& \dfrac{1}{4(2\pi)^{16}A^{+2}B^{-2}} \sum_{s_it_is'_it'_i}\int d\bar{A}^+ d\bar{B}^- dJ_1^+dJ_1^- \hat{\sigma}^{s_1,t_1;\tilde{s}_1,\tilde{t}_1;\mu_1}_{q\bar{q} \to \gamma*}(\hat{s} = 2J_1^+J_1^-)
\\ \nonumber
&\times \hat{\sigma}^{s_2,t_2;\tilde{s}_2,\tilde{t}_2;\mu_2}_{q\bar{q} \to \gamma*}(\hat{s} = 2J_2^+J_2^-)   \int d^2\vect{a_1} d^2\vect{b_1} d^2\vect{\Delta} d^2\vect{\bar{A}} d\bar{A}^- d^2\vect{\bar{B}} d\bar{B}^+
\\ \nonumber
&\times \sum_\chi \psi_{p;q\bar{q}}^{s_1s_2\chi}(J_1^+,J_2^+,\vect{a_1},\vect{a_2},\bar{A}^-) \psi_{p;q\bar{q}}^{*\tilde{s}_1\tilde{s}_2\chi}(J_1^{+},J_2^{+},\vect{a_1+\Delta},\vect{a_2-\Delta},\bar{A}^-)4J_1^+J_2^+A^+
\\ \nonumber
&\times  \sum_\gamma \psi_{p;q\bar{q}}^{t_2t_1\gamma}(J_2^-,J_1^-,\vect{b_2},\vect{b_1},\bar{B}^+)\psi_{p;q\bar{q}}^{*\tilde{t}_2\tilde{t}_1\gamma}(J_2^{-},J_1^{-},\vect{b_2+\Delta},\vect{b_1-\Delta},\bar{B}^+)4J_1^-J_2^-B^-
\end{align}

In the second line of \eqref{2v2XS2}, we have introduced the transverse variable $\vect{\Delta}$ via $\vect{a_1} = \vect{\tilde{a}_1+ \Delta}$,
and converted the integral over $\vect{J_1}$ to an integral over $\vect{b_1}$ using $\vect{a_1+b_1} = \vect{J_1}$. We have also made use
of the fact that $s = 2A^+B^-$. Let us define the `nonperturbatively generated parton pair' $\vect{r}$-space 2pGPD according to:
\begin{align} \label{2pGPDkspace}
 \Gamma_{p;q\bar{q}}^{s_1s_2,\tilde{s}_1\tilde{s}_2}\left(\dfrac{J_1^+}{A^+},\dfrac{J_2^+}{A^+};\vect{\Delta}\right) \equiv&  \dfrac{2}{(2\pi)^7}
\sum_\chi\int  d\bar{A}^- d^2\vect{\bar{A}}d^2\vect{a_1} \psi_{p;q\bar{q}}^{s_1s_2\chi}(J_1^+,J_2^+,\vect{a_1},\vect{a_2},\bar{A}^-,\vect{\bar{A}})
\\ \nonumber
& \times \psi_{p;q\bar{q}}^{*\tilde{s}_1\tilde{s}_2\chi}(J_1^+,J_2^+,\vect{a_1+\Delta},\vect{a_2-\Delta},\bar{A}^-,\vect{\bar{A}}) J_1^+J_2^+A^+
\end{align}
We also introduce the following scaling variables:
\begin{align} \label{scalingvars}
 x_1 \equiv J_1^+/A^+ \qquad x_2 \equiv J_2^+/A^+ \qquad y_1 \equiv J_1^-/B^- \qquad y_2 \equiv J_2^-/B^-
\end{align}

Changing variables in \eqref{2v2XS2} to the scaling variables, replacing appropriate combinations of $\psi$s by
$\Gamma$s according to \eqref{2pGPDkspace}, and using the obvious relation $\Gamma_{p;\bar{q}q}^{s_1s_2,\tilde{s}_1\tilde{s}_2}(x_1,x_2;\vect{\Delta}) =
\Gamma_{p;q\bar{q}}^{s_2s_1,\tilde{s}_2\tilde{s}_1}(x_2,x_1;-\vect{\Delta})$, we finally obtain:
\begin{align}
\sigma_{2v2}(s)& =  \sum_{s_it_i\tilde{s}_i\tilde{t}_i}\int dx_1dx_2dy_1dy_2 \hat{\sigma}^{s_1,t_1;\tilde{s}_1,\tilde{t}_1;\mu_1}_{q\bar{q} \to \gamma*}(\hat{s} = x_1y_1s)
\hat{\sigma}^{s_2,t_2;\tilde{s}_2,\tilde{t}_2;\mu_2}_{q\bar{q} \to \gamma*}(\hat{s} = x_2y_2s)
\\ \nonumber
&\times   \int \dfrac{d^2\vect{\Delta}}{(2\pi)^2}\Gamma_{p;q\bar{q}}^{s_1s_2,\tilde{s}_1\tilde{s}_2}(x_1,x_2;\vect{\Delta}) \Gamma_{p;\bar{q}q}^{t_1t_2,\tilde{t}_1\tilde{t}_2}(y_1,y_2;\vect{-\Delta})
\end{align}

The cross section is of the anticipated form \eqref{2v2XS}. The most important result of this preliminary calculation is the
definition of the `nonperturbatively generated parton pair' $\vect{r}$-space 2pGPD \eqref{2pGPDkspace}, which we shall make use of later.

The calculation of the cross section contribution associated with figure \ref{fig:DPSdiagrams}(a), 
$\sigma_{2v1}(s)$, proceeds in a very similar manner to the calculation of $\sigma_{2v2}(s)$. Once 
again we work in a frame in which $A = A^+p$ and $B = B^+ n$. We can directly write down the following 
expression for the cross section:
\begin{align} \label{2v1XS}
\sigma_{2v1}(s) =& \dfrac{1}{2(2\pi)^6s} \sum_{\chi}\int d^4\bar{A} d^4J_1d^4J_2 \delta(J_1^2-Q^2)\delta(J_2^2-Q^2)\delta^{(4)}(\bar{A}+J_1+J_2-A-B)
\\ \nonumber
&\times \mathcal{M}^{\lambda;\chi\mu_1\mu_2}(A,B;\bar{A},J_1,J_2) \mathcal{M}^{\lambda;\chi\mu_1\mu_2}(A,B;\bar{A},J_1,J_2)^*
\end{align}
where:
\begin{align} \label{2v1ME}
\mathcal{M}^{\lambda;\chi\mu_1\mu_2}&(A,B;\bar{A},J_1,J_2)
\\ \nonumber
\equiv& \int \dfrac{d^4a_1}{(2\pi)^4} i^2Tr(\slashed{a}_1  \varphi^\chi_p(a_1,a_2,\bar{A}) \slashed{a}_2 T^{\lambda;\mu_1\mu_2}(a_2a_1B \to J_1J_2))/[D(a_1)D(a_2)],
\end{align}
\begin{align}
T^{\lambda;\mu_1\mu_2}(a_2a_1B \to J_1J_2) \equiv i^5(eQ_q)^2g_s\dfrac{\slashed{\varepsilon}_{\mu_2}^*(J_2)\slashed{b}_2
\slashed{\varepsilon}_{\lambda}(B)\slashed{b}_1\slashed{\varepsilon}_{\mu_1}^*(J_1)}{D(b_1)D(b_2)}
\end{align}

The lines with momentum $a_1$ are restricted to small virtuality by $\varphi_A$, so we can decompose the slashed $a_i$ vectors in
\eqref{2v1ME} into outer products of particle or antiparticle spinors:
\begin{align} \label{2v1ME2}
\mathcal{M}^{\lambda;\chi\mu_1\mu_2}&(A,B;\bar{A},J_1,J_2)
 \\ \nonumber
\simeq & \sum_{s_i}\int \dfrac{d^4a_1}{(2\pi)^4} \left[-\dfrac{\bar{u}^{s_1}(a_1) \varphi^\chi_p(a_1,a_2,\bar{A}) v^{s_2}(a_2)}{D(a_1)D(a_2)} \right]
 \mathcal{M}^{s_2s_1\lambda;\mu_1\mu_2}(a_2a_1B \to J_1J_2)
\end{align}
$\mathcal{M}^{s_2s_1\lambda;\mu_1\mu_2}(a_2a_1B \to J_1J_2)$ is the matrix element for $q\bar{q}g \to \gamma^*\gamma^*$ with initial quark and
antiquark having small transverse momentum and virtuality.

For reasons similar to those leading to equation \eqref{2v2ME3}, we can move the $a_1^-$ integration such that it only acts on the
part of \eqref{2v1ME2} in square brackets, and set $a_1^-=0$ in the rest of the integrand. Provided that that $\vect{J_1}^2 \gg \Lambda^2$,
 we can perform an analogous operation for the $\vect{a_1}$ integration. The reason for
this is that when $\vect{J_1}^2 \gg \Lambda^2$, the transverse momenta of the $a_i$ lines (constrained to be of order $\Lambda$ by
$\varphi_A$) are negligible compared to the transverse momenta of the $b_i$ and $J_i$ lines in
$\mathcal{M}^{s_2s_1\lambda;\mu_1\mu_2}(a_2a_1B \to J_1J_2)$, so we make only a small error by setting $\vect{a_i}$ to zero in this
factor provided $\vect{J_1}^2 \gg \Lambda^2$. Applying these approximations:
\begin{align} \label{2v1ME3}
\mathcal{M}^{\lambda;\chi\mu_1\mu_2}&(A,B;\bar{A},J_1,J_2)
 \\ \nonumber
\simeq & \sum_{s_i} \int \dfrac{d a_1^+}{2\pi} \mathcal{M}^{s_2s_1\lambda;\mu_1\mu_2}(a_2a_1B \to J_1J_2) |_{a_1^-= 0,\vect{a_1}=0}
\\ \nonumber
&\times  -\int \dfrac{d^2\vect{a_1}}{(2\pi)^2} \dfrac{d a_1^-}{2\pi} \dfrac{[\bar{u}^{s_1}(a_1) \varphi^\chi_p(a_1,a_2,\bar{A}) v^{s_2}(a_2)]}{D(a_1)D(a_2)}
\end{align}

We identify the final factor in \eqref{2v1ME3} as the integral of $\psi_p$ over $\vect{a_1}$. Writing out the denominator factors
in $\mathcal{M}^{s_2s_1\lambda;\mu_1\mu_2}(a_2a_1B \to J_1J_2)$ explicitly we have:
\begin{align} \label{2v1ME4}
\mathcal{M}^{\lambda;\chi\mu_1\mu_2}&(A,B;\bar{A},J_1,J_2)
\\ \nonumber
\simeq & \sum_{s_i} \int \dfrac{d a_1^+}{2\pi} \left[ \int d^2\vect{a_1}/(2\pi)^2\psi_p^{s_1s_2\chi}(a_1^+,a_2^+,\vect{a_1},\vect{a_2},\bar{A}^-) \right]
\\ \nonumber
& \times \dfrac{\mathcal{T}^{s_2s_1\lambda;\mu_1\mu_2}(a_2a_1B \to J_1J_2) |_{a_1^-= 0,\vect{a_1}= 0}}{[2(J_1^+-a_1^+)J_1^--\vect{J_1}^2+i\epsilon]
[2(a_1^+-J_1^+)J_2^--\vect{J_1}^2+i\epsilon]}
\end{align}
where:
\begin{align}
\mathcal{T}^{s_2s_1\lambda;\mu_1\mu_2}(a_2a_1B \to J_1J_2) \equiv i^5g_s(eQ_q)^2\bar{v}^{s_2}(a_2)\slashed{\varepsilon}_{\mu_2}^*(J_2)\slashed{b}_2
\slashed{\varepsilon}_{\lambda}(B)\slashed{b}_1\slashed{\varepsilon}_{\mu_1}^*(J_1) {u}^{s_1}(a_1)
\end{align}

Examination of the denominator factors in \eqref{2v1ME3} reveals that the majority of the contribution to the $a_1^+$ integration comes
from the region $a_1^+ \sim J_1^+$. For this reason we can set $a_1^+ = J_1^+$ in the numerator before evaluating the $a_1^+$ integral
using contour integration:
\begin{align} \label{2v1ME5}
\mathcal{M}^{\lambda;\chi\mu_1\mu_2}&(A,B;\bar{A},J_1,J_2)
\\ \nonumber
\simeq & \sum_{s_i} {i \left[ \int d^2\vect{a_1}/(2\pi)^2\psi_p^{s_1s_2\chi}(J_1^+,J_2^+,\vect{a_1},\vect{a_2},\bar{A}^-) \right]
}
\\ \nonumber
& \times \dfrac{\mathcal{T}^{s_2s_1\lambda;\mu_1\mu_2}(a_2a_1B \to J_1J_2) |_{a_1^-= 0,\vect{a_1}= 0,a_1^+=J_1^+}}{2(J_1^-+J_2^-)\vect{J_1}^2}
\end{align}

We are interested in the behaviour of $\mathcal{M}^{\lambda;\chi\mu_1\mu_2}(A,B;\bar{A},J_1,J_2)$ when
$\vect{J_1}^2 \ll Q^2$ (but still $\gg \Lambda^2$) such that all of the internal particles have
transverse momenta and virtualities much less than $Q$. In this limit we can use spinor completeness
relations to split $\mathcal{T}$ up into two $q\bar{q} \to \gamma^*$ matrix elements and one $g \to q\bar{q}$
matrix element, with the quark and antiquark having small transverse momenta and virtuality
$\mathcal{O}(|\vect{J_1}|)$ in each matrix element. The quark and antiquark transverse momenta and
virtualities can be set to zero in the `hard' $q\bar{q} \to \gamma^*$ matrix elements with only a small
accompanying error $\mathcal{O}(\vect{J_1}^2/Q^2)$, but we must keep the term proportional to
$\vect{J_1}$ in the $g \to q\bar{q}$ matrix element as this vanishes in the limit $\vect{J_1} \to 0$:
\begin{align} \label{2v1ME6}
\mathcal{M}^{\lambda;\chi\mu_1\mu_2}&(A,B;\bar{A},J_1,J_2)
\\ \nonumber
\simeq & \sum_{s_it_i} \dfrac{-i \left[ \int d^2\vect{a_1}/(2\pi)^2\psi_p^{s_1s_2\chi}(a_1^+,a_2^+,\vect{a_1},\vect{a_2},\bar{A}^-) \right] \mathcal{M}^{\lambda \to t_1t_2}_{g \to \bar{q}q}(B;J_1^-n+\vect{J}_1,J_2^-n+\vect{J}_2)}
{2(J_1^-+J_2^-)\vect{J_1}^2}
\\ \nonumber
& \times \mathcal{M}^{t_1s_1 \to \mu_1}_{\bar{q}q \to \gamma^*}(J_1^-n,J_1^+p;J_1^-n+J_1^+p)\mathcal{M}^{t_2s_2 \to \mu_2}_{q\bar{q} \to \gamma^*}(J_2^-n,J_2^+p;J_2^-n+J_2^+p)
\end{align}

Having inserted \eqref{2v1ME6} into \eqref{2v1XS}, we use \eqref{hardXS} and the following connection between
$\mathcal{M}_{g \to \bar{q}q}$ and helicity-dependent unregularised splitting functions in the result \cite{Peskin:1995ev}:
\begin{align} \label{1->2sptfn}
\dfrac{J_1^-J_2^-}{(J_1^-+J_2^-)^2} &\mathcal{M}^{\lambda \to t_1t_2}_{g \to \bar{q}q}(B;J_1^-n+\vect{J}_1,J_2^-n+\vect{J}_2)
\mathcal{M}^{*\lambda \to \tilde{t}_1\tilde{t}_2}_{g \to \bar{q}q}(B;J_1^-n+\vect{J}_1,J_2^-n+\vect{J}_2)
\\ \nonumber
&= 2g_s^2P_{g \to q\bar{q}}^{\lambda \to t_2t_1,\tilde{t}_2\tilde{t}_1}\left(\dfrac{J_2^-}{J_1^-+J_2^-}\right) \vect{J_1}^2
\end{align}

This yields:
\begin{align}\nonumber
\sigma_{2v1}(s) =& \sum_{s_i\tilde{s}_it_i\tilde{s}_i\chi}\dfrac{4}{(2\pi)^{12}s}\int d^4\bar{A}  d^4J_1
\hat{\sigma}^{s_1,t_1;\tilde{s}_1,\tilde{t}_1;\mu_1}_{\bar{q}q \to \gamma*}(\hat{s} = 2J_1^+J_1^-) J_1^+
\hat{\sigma}^{s_2,t_2;\tilde{s}_2,\tilde{t}_2;\mu_2}_{q\bar{q} \to \gamma*}(\hat{s} = 2J_2^+J_2^-) J_2^+
\\  \label{2v1XS2}
& \times   \left[ \int d^2\vect{a_1}\psi_p^{s_1s_2\chi}(J_1^+,J_2^+,\vect{a_1},\vect{a_2},\bar{A}^-) \right]
\\ \nonumber
& \times \left[ \int d^2\vect{a'_1}\psi_p^{*\tilde{s}_1\tilde{s}_2\chi}(J_1^+,J_2^+,\vect{\tilde{a}_1},\vect{\tilde{a}_2},\bar{A}^-) \right]
g_s^2P_{g \to q\bar{q}}^{\lambda \to t_2t_1,\tilde{t}_2\tilde{t}_1}\left(\dfrac{J_2^-}{J_1^-+J_2^-}\right) \dfrac{1}{\vect{J_1}^2}
\displaybreak[3]
\\ \nonumber
=& \sum_{s_i\tilde{s}_it_i\tilde{t}_i}\dfrac{1}{(2\pi)^{3}A^{+2}B^-}\int   d^4J_1 d\bar{A}^+
\hat{\sigma}^{s_1,t_1;\tilde{s}_1,\tilde{t}_1;\mu_1}_{\bar{q}q \to \gamma*}(\hat{s} = 2J_1^+J_1^-)
\hat{\sigma}^{s_2,t_2;\tilde{s}_2,\tilde{t}_2;\mu_2}_{q\bar{q} \to \gamma*}(\hat{s} = 2J_2^+J_2^-)
\\  \label{2v1XS3}
& \times
\biggl[ \dfrac{2}{(2\pi)^7}\sum_{\chi}\int \dfrac{d^2\vect{\Delta}}{(2\pi)^2}d^2\vect{\bar{A}}d\bar{A}^- d^2\vect{a_1}  \psi_p^{s_1s_2\chi}(J_1^+,J_2^+,\vect{a_1},\vect{a_2},\bar{A}^-)
\\ \nonumber
& \times
\psi_p^{*\tilde{s}_1\tilde{s}_2\chi}(J_1^+,J_2^+,\vect{a_1+\Delta},\vect{a_2-\Delta},\bar{A}^-) J_1^+ J_2^+ A^+ \biggr]
g_s^2P_{g \to q\bar{q}}^{\lambda \to t_2t_1,\tilde{t}_2\tilde{t}_1}\left(\dfrac{J_2^-}{J_1^-+J_2^-}\right) \dfrac{1}{\vect{J_1}^2}
\end{align}

In \eqref{2v1XS3} we have once again introduced the transverse variable $\vect{\Delta}$ via the same relation as in the
$2v2$ case. We recognise the object in square brackets in \eqref{2v1XS3} as the integral of the nonperturbatively generated parton pair
$\vect{r}$-space 2pGPD over $\vect{\Delta}$. If we make a change of longitudinal integration variables in \eqref{2v1XS3}
to the scaling variables \eqref{scalingvars}, then we finally obtain:
\begin{align} \label{2v1XS4}
\sigma_{2v1}(s) =& \sum_{s_i\tilde{s}_it_i\tilde{t}_i}\int dx_1dx_2dy_1dy_2
 \hat{\sigma}^{s_1,t_1;\tilde{s}_1,\tilde{t}_1;\mu_1}_{\bar{q}q \to \gamma*}(\hat{s} = x_1y_1s)
\hat{\sigma}^{s_2,t_2;\tilde{s}_2,\tilde{t}_2;\mu_2}_{q\bar{q} \to \gamma*}(\hat{s} = x_2y_2s)
\\  \nonumber
& \times \left[\int \dfrac{d^2\vect{\Delta}}{(2\pi)^2} \Gamma_{p;q\bar{q}}^{s_1s_2,\tilde{s}_1\tilde{s}_2}\left(x_1,x_2;\vect{\Delta}\right)\right]\left[\dfrac{\alpha_s}{2\pi}P_{g \to q\bar{q}}^{\lambda \to t_2t_1,\tilde{t}_2\tilde{t}_1}\left(y_2\right)\delta(1-y_1-y_2) \int_{\Lambda^2}^{Q^2} \dfrac{d\vect{J_1}^2}{\vect{J_1}^2}\right]
\end{align}

We have restricted our integration over $\vect{J_1}^2$ to the range ${\Lambda^2}<\vect{J_1}^2< {Q^2}$, which corresponds
to the range over which our approximate expression for the matrix element \eqref{2v1ME6} is valid. The contributions to
$\sigma_{2v1}$ coming from $\vect{J_1}^2$ values outside this range do not have the same $1/\vect{J_1}^2$ structure.

The integral over $\vect{J_1}$ in \eqref{2v1XS4} gives rise to a large transverse momentum logarithm $\log(Q^2/\Lambda^2)$, whilst the
integral over $\vect{\Delta}$ gives a prefactor of order $\Lambda^2 \sim 1/R_p^2$ (since the nonperturbatively generated parton pair
$\vect{r}$-space 2pGPD only has support for transverse momenta, and therefore transverse momentum imbalances $\vect{r}$, of order $\Lambda_{QCD}$). Thus, as we asserted at the
beginning of this section, there is a part of the cross section expression for figure \ref{fig:DPSdiagrams}(a) that is
proportional to $\log(Q^2/\Lambda^2)/R_p^2$ and should be included in the LO DPS cross section.

Note that the quantity $\int d^2\vect{\Delta}\Gamma_p^{s_1s_2,\tilde{s}_1\tilde{s}_2}\left(x_1,x_2;\vect{\Delta}\right)
/(2\pi)^2$ is equal to the $\vect{b}$-space nonperturbatively generated parton pair 2pGPD evaluated at zero transverse separation, \\
$\Gamma_p^{s_1s_2,\tilde{s}_1\tilde{s}_2}\left(x_1,x_2;\vect{b=0}\right)$. This appears to indicate that the 2v1 
contribution to DPS probes nonperturbatively generated parton pair 2pGPDs at zero parton separation. In fact, the result \eqref{2v1XS4} 
actually corresponds to a broad logarithmic integral over values of $\vect{b}^2$ that are $\ll R_p^2$ but $\gg 1/Q^2$. 
The $\vect{b}$-space 2pGPD evaluated at $\vect{b=0}$ appears in \eqref{2v1XS4} because the $\vect{r}$-space nonperturbatively 
generated parton pair 2pGPD dies off rapidly for $\vect{\Delta}^2 \gg \Lambda^2$, which is equivalent to the $\vect{b}$-space 
nonperturbatively generated parton pair 2pGPD not containing any fluctuations with length scales $\ll R_p$. Then we can approximate 
$\Gamma_p(\vect{b})$ for the relevant values of $\vect{b}$ in \eqref{2v1XS4} by $\Gamma_p(\vect{b=0})$.

If one assumes that diagrams of the form of figure \ref{fig:2v21v2}(b) are the only diagrams of the `2v1' type that
contribute to the DPS cross section at leading logarithmic order, then a generalisation of the result in
\eqref{2v1XS4} yields the expression below for the contribution of 2v1 graphs to the LO DPS cross 
section\footnote{Note that here and in the rest of this section we will take the scales associated 
with the two hard scales to be equal, $Q_A^2 = Q_B^2 = Q^2$. We will comment in section \ref{sec:XsecResults} on 
the generalisation of the results of this section to the case of unequal scales. Note also that we only 
write down the unpolarised diagonal contribution in colour, flavour and spin space here. The contributions 
associated with spin polarisation (either longitudinal or transverse) and flavour interference are expected 
to have a similar structure. On the other hand, it is known that the colour correlation/interference and
parton type interference contributions will be suppressed by Sudakov factors, as is discussed in
\cite{Mekhfi:1988kj, Diehl:2011tt, Diehl:2011yj, Manohar:2012jr}.}:
\begin{align} \label{1v2XSec}
\sigma^{D,2v1}_{(A,B)}(s) =& 2\times\dfrac{m}{2} \sum_{l i_i j_i i'_i j'_i} \int_{\Lambda^2}^{Q^2} dk^2 \dfrac{\alpha_s\left(k^2\right)}{2\pi k^2} \int dx_1 dx_2 dy_1 dy_2 
\dfrac{dx'_1}{x'_1} \dfrac{dx'_2}{x'_2} \dfrac{dy'_1}{y'_1} \dfrac{dy'_2}{y'_2} 
\\ \nonumber
& \times \hat{\sigma}_{i_1j_1 \to A}(\hat{s} = x_1y_1s)
\hat{\sigma}_{i_2j_2 \to B}(\hat{s} = x_2y_2s)
\\ \nonumber
& \times \dfrac{D^{l}_p(y'_1+y'_2,k^2)}{y'_1+y'_2} P_{l \to j_1'j_2'}\left(\dfrac{y'_1}{y'_1+y'_2}\right)
D_{j_1'}^{j_1}\left(\dfrac{y_1}{y'_1};k^2,Q^2\right) D_{j_2'}^{j_2}\left(\dfrac{y_2}{y'_2};k^2,Q^2\right)
\\ \nonumber
& \times D_{i_1'}^{i_1}\left(\dfrac{x_1}{x'_1};\Lambda^2,Q^2\right) D_{i_2'}^{i_2}\left(\dfrac{x_2}{x'_2};
\Lambda^2,Q^2\right) \Gamma^{i_1'i_2'}_{p,indep}(x'_1, x'_2, \vect{b} = \vect{0}; \Lambda^2)
\end{align}

$D_{i}^{j} \left(x;k^2,Q^2\right)$ are the Green's functions of the DGLAP equations -- i.e. a set of 
functions obeying the DGLAP equations with the initial condition $D_{i}^{j} \left(x;k^2,k^2\right) = 
\delta_{ij} \; \delta(1-x)$. $\Gamma^{i_1'i_2'}_{p,indep}(x'_1, x'_2; \vect{b} = \vect{0}, \Lambda^2)$
represents a nonperturbative initial condition for the two independent ladders in figure \ref{fig:2v21v2}(b).
In \eqref{1v2XSec} we have re-inserted the symmetry factor $m/2$ that has been omitted in earlier 
discussion in this section ($m=1$ if the two hard processes are identical, and $m=2$ otherwise). 
There is an additional prefactor of $2$ in \eqref{1v2XSec} because there are 
two sets of 2v1 graphs that give equivalent contributions -- in one set the nonperturbatively generated 
parton pair emerges from the `left' proton, whilst in the other it emerges from the `right' proton.

Equation \eqref{1v2XSec} can be written in a more compact fashion as:
\begin{align} \label{1v2XSec2}
\sigma^{D,2v1}_{(A,B)}(s) = & 2\times\dfrac{m}{2}\sum_{i_ij_i}\int dx_1dx_2dy_1dy_2 \hat{\sigma}_{i_1j_1 \to A}(\hat{s} = x_1y_1s)\hat{\sigma}_{i_2j_2 \to B}(\hat{s} = x_2y_2s)
\\ \nonumber
&\times \breve{D}^{j_1j_2}_p(y_1,y_2;Q^2)  \int \dfrac{d^2\vect{\Delta}}{(2\pi)^2} \Gamma^{i_1i_2}_{p,indep}(x_1,x_2,\vect{\Delta};Q^2)
\\ \nonumber
 = & 2\times\dfrac{m}{2} \sum_{i_ij_i} \int dx_1dx_2dy_1dy_2 \hat{\sigma}_{i_1j_1 \to A}(\hat{s} = x_1y_1s)\hat{\sigma}_{i_2j_2 \to B}(\hat{s} = x_2y_2s)
\\ \nonumber
&\times \breve{D}^{j_1j_2}_p(y_1,y_2;Q^2)  \Gamma^{i_1i_2}_{p,indep}(x_1,x_2,\vect{b=0};Q^2)
\end{align}
where:
\begin{align} \label{indepbranchGPD}
\Gamma^{i_1i_2}_{p,indep}(x_1,x_2,\vect{b};Q^2) \equiv& \sum_{i'_i} \int  \dfrac{dx'_1}{x'_1} \dfrac{dx'_2}{x'_2} D_{i_1'}^{i_1}\left(\dfrac{x_1}{x'_1};\Lambda^2,Q^2\right) D_{i_2'}^{i_2}\left(\dfrac{x_2}{x'_2};
\Lambda^2,Q^2\right) 
\\ \nonumber
& \times \Gamma^{i_1'i_2'}_{p,indep}(x'_1, x'_2, \vect{b}; \Lambda^2)
\\ \label{Dbreve}
\breve{D}^{j_1j_2}_p(y_1,y_2;Q^2) \equiv& \sum_{lj'_i} \int_{\Lambda^2}^{Q^2} dk^2 \dfrac{\alpha_s\left(k^2\right)}{2\pi k^2} \dfrac{dy'_1}{y'_1} \dfrac{dy'_2}{y'_2}\dfrac{D^{l}_p(y'_1+y'_2,k^2)}{y'_1+y'_2} 
\\ \nonumber
&\times P_{l \to j_1'j_2'}\left(\dfrac{y'_1}{y'_1+y'_2}\right)
D_{j_1'}^{j_1}\left(\dfrac{y_1}{y'_1};k^2,Q^2\right) D_{j_2'}^{j_2}\left(\dfrac{y_2}{y'_2};k^2,Q^2\right)
\end{align}

As mentioned in section \ref{sec:Xsecintro}, and as will be explored in detail in section \ref{sec:crosstalk}, there
are additional diagrams of the `2v1' type that contribute at leading logarithmic order to the DPS cross 
section, aside from those represented by figure \ref{fig:2v21v2}(b). These involve crosstalk interactions 
between the two nonperturbatively generated ladders. Equation \eqref{1v2XSec} (or \eqref{1v2XSec2}) therefore 
represents only part of the 2v1 contribution to the LO DPS cross section. For the moment, however, we'll limit 
our discussion to just this part.

A necessary requirement for \eqref{1v2XSec} (or \eqref{1v2XSec2}) to be valid (at least as an incomplete part of a contribution 
to the DPS cross section) is that the independent two-ladder 2pGPD $\Gamma^{i_1i_2}_{p,indep}
(x_1, x_2; \vect{b}, Q^2)$ should be smooth on distance scales $\ll R_p \sim 1/\Lambda$ (or 
equivalently that the corresponding distribution in terms of the transverse momentum imbalance $\vect{\Delta}$
is cut off at values of order $\Lambda$). This appears to be a somewhat reasonable requirement -- at the scale
$\Lambda$ there is only this scale available to set the size of the $\vect{\Delta}$ profile for $\Gamma^{i_1i_2}
_{p,indep}(x_1, x_2; \vect{\Delta}, \Lambda^2)$, and the evolution equation for the independent two-ladder 2pGPD 
(which is just the `double DGLAP' equation of \cite{Shelest:1982dg, Zinovev:1982be} with the `single PDF feed
term' defined in \cite{Gaunt:2009re} removed) preserves the transverse profile. In any case, such behaviour 
for $\Gamma^{i_1i_2}_{p,indep}(x_1, x_2; \vect{\Delta}, Q^2)$ would appear to be required in order to get 
the necessary prefactor of order $1/R_p^2$ in the 2v2 contribution to DPS, which is calculated according 
to the following expression (for the diagonal unpolarised contribution):
\begin{align} \label{2v2XSec}
\sigma^{D,2v2}_{(A,B)}(s) = & \dfrac{m}{2} \sum_{i_ij_i}\int dx_1dx_2dy_1dy_2 \hat{\sigma}_{i_1j_1 \to A}(\hat{s} = x_1y_1s)\hat{\sigma}_{i_2j_2 \to B}(\hat{s} = x_2y_2s)
\\ \nonumber
&\times \int \dfrac{d^2\vect{\Delta}}{(2\pi)^2} \Gamma^{i_1i_2}_{p,indep}(x_1,x_2,\vect{\Delta};Q^2) \Gamma^{j_1j_2}_{p,indep}(y_1,y_2,\vect{-\Delta};Q^2)
\displaybreak[3] \\ \nonumber
 = & \dfrac{m}{2} \sum_{i_ij_i} \int dx_1dx_2dy_1dy_2 \hat{\sigma}_{i_1j_1  \to A}(\hat{s} = x_1y_1s)\hat{\sigma}_{i_2j_2 \to B}(\hat{s} = x_2y_2s)
\\ \nonumber
&\times \int d^2\vect{b} \Gamma^{i_1i_2}_{p,indep}(x_1,x_2,\vect{b};Q^2) \Gamma^{j_1j_2}_{p,indep}(y_1,y_2,\vect{b};Q^2)
\end{align}

If one assumes that $\Gamma^{ij}_{p,indep}(x_1, x_2; \vect{b}, Q^2)$ can be factorised into
a longitudinal piece $\tilde{D}_{p,indep}^{ij}(x_1,x_2;Q^2)$ and a flavour-independent transverse piece
$F(\vect{b})$, where  $F(\vect{b})$ is a smooth function of radius $R_p$ normalised to $1$, then 
\eqref{1v2XSec} and \eqref{2v2XSec} become:
\begin{align} \label{1and2XS2a}
\sigma^{D,2v2}_{(A,B)}(s) = & \dfrac{m}{2} \sum_{i_ij_i} \dfrac{1}{\sigma_{eff,2v2}} \int dx_1dx_2dy_1dy_2 \hat{\sigma}_{i_1j_1 \to A}(\hat{s} = x_1y_1s)
\\ \nonumber
&\times \hat{\sigma}_{i_2j_2 \to B}(\hat{s} = x_2y_2s) \tilde{D}^{i_1i_2}_{p,indep}(x_1,x_2;Q^2) \tilde{D}^{j_1j_2}_{p,indep}(y_1,y_2;Q^2)
\\ \label{1and2XS2b}
\sigma^{D,2v1}_{(A,B)}(s) = & 2\times\dfrac{m}{2} \sum_{i_ij_i} \dfrac{1}{\sigma_{eff,2v1}} \int dx_1dx_2dy_1dy_2 \hat{\sigma}_{i_1j_1 \to A}(\hat{s} = x_1y_1s)
\\ \nonumber
&\times \hat{\sigma}_{i_2j_2 \to B}(\hat{s} = x_2y_2s) \breve{D}^{j_1j_2}_p(y_1,y_2;Q^2)  \tilde{D}^{i_1i_2}_{p,indep}(x_1,x_2;Q^2)
\end{align}
where:
\begin{align}
\dfrac{1}{\sigma_{eff,2v2}} \equiv \int d^2\vect{b} [F(\vect{b})]^2 = \int \dfrac{d^2\vect{\Delta}}{(2\pi)^2} [F(\vect{\Delta})]^2
\\
\dfrac{1}{\sigma_{eff,2v1}} \equiv  F(\vect{b}=\vect{0}) = \int \dfrac{d^2\vect{\Delta}}{(2\pi)^2} [F(\vect{\Delta})]
\end{align}

$F(\vect{\Delta})$ is the Fourier transform of $F(\vect{b})$. We see that the geometrical prefactors
for the two different contributions to the DPS cross section are different in general, $\sigma_{eff,2v2} 
\neq \sigma_{eff,2v1}$. If one assumes that two nonperturbatively generated ladders
are to some degree uncorrelated in transverse space, $F(\vect{b})$ is given by a convolution of an 
azimuthally symmetric transverse parton density in the proton $\rho(\vect{r})$ with itself, where 
$\rho(\vect{r})$ must be normalised to $1$ in order to ensure the appropriate normalisation of 
$F(\vect{b})$:
\begin{align} \label{F&rho}
F(\vect{b}) = \int d^2\vect{r} \rho(\vect{r}) \rho(\vect{b-r})
\end{align}

Then, if one takes the Gaussian form $\exp[-r^2/(2R^2)]/(\pi R^2)$ for $\rho$ (with $R$ a constant parameter), 
one finds that $\sigma_{eff,2v1} = \sigma_{eff,2v2}/2$ -- that is, the 2v1 contribution receives a factor of $2$ 
enhancement over the 2v2 contribution from the geometrical prefactor alone (in the next section, we'll discover 
that the 2v1 contribution is further enhanced at low $x$ as a result of the crosstalk interactions on the 
two-ladder side that are allowed for this contribution). The ratio $\sigma_{eff,2v2}/\sigma_{eff,2v1}$ does not 
depend much on the precise shape of $\rho$ -- for example, one obtains 2.18 if $\rho$ is a top hat 
$\tfrac{1}{\pi R^2} \Theta(R-r)$, 2.32 if $\rho$ is the projection of an exponential $\int dz \tfrac{1}{8\pi R^3} 
\exp(\sqrt{-r^2+z^2}/R)$, and 1.94 if $\rho$ is the projection of a hard sphere $\tfrac{3}{2\pi R^2}(1-r^2/R^2)
^{1/2}\Theta(R-r)$ (with $R$ once again a constant parameter in these expressions). This `factor of two' 
enhancement of each 2v1 contribution over the 2v2 contribution from the geometrical prefactor has previously
been noted in \cite{Blok:2011bu, Blok:2012mw}. It is important to bear in mind, however, that in order to 
obtain an enhancement that is roughly a factor of $2$ one has to make a number of assumptions whose validity 
is somewhat uncertain (this is particularly the case for the assumption \eqref{F&rho}). There could be some 
`clustering' of the nonperturbative partons in transverse space, which would tend to increase 
$\sigma_{eff,2v2}/\sigma_{eff,2v1}$. Alternatively it is not inconceivable that the probability to find two 
nonperturbative partons separated by small distances $\ll R_p$ could be smaller than the probability to find 
them separated by distances of order $R_p$ -- in this scenario $\sigma_{eff,2v2}/\sigma_{eff,2v1}$ would be 
reduced.

\section{Crosstalk between Ladders in the 2v1 Contribution} \label{sec:crosstalk}

\begin{figure}
\centering
\includegraphics[scale=0.6]{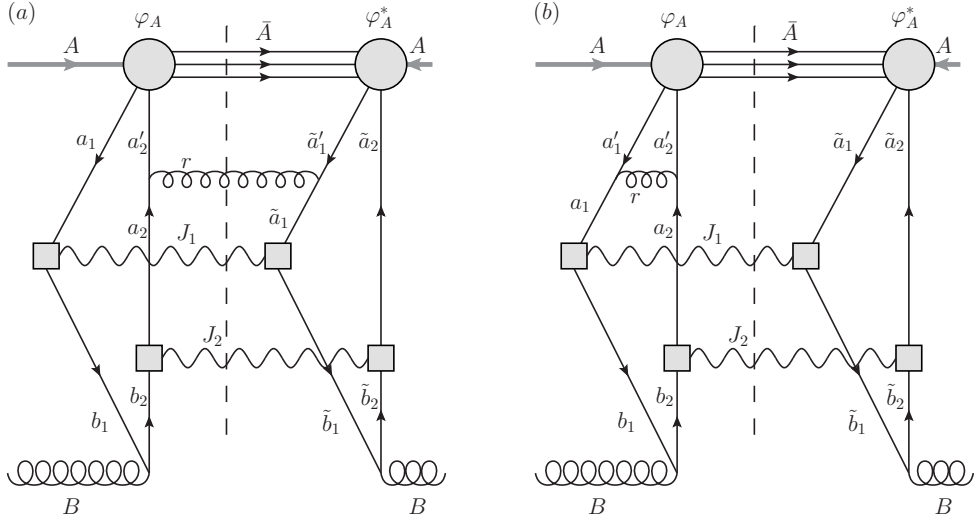}
\caption{(a) Simple 2v1 diagram including an `off-diagonal real emission' process. (b) Simple 2v1 diagram including 
a `virtual exchange' process.}
\label{fig:recombDPS}
\end{figure}

In the previous section we demonstrated that there is a leading logarithmic contribution to the DPS cross section 
associated with diagrams in which a single parton ladder from one proton splits into two, and then the two daughter
ladders interact with two independent ladders from the other proton (that are only connected to one another via 
low-scale nonperturbative interactions). It is suggested in a number of works \cite{Ryskin:2011kk, Blok:2011bu, 
Ryskin:2012qx} that these diagrams are the only ones involving a single $1 \to 2$ ladder branching that give rise
to a leading logarithmic contribution to DPS. Here, we show that there is also a leading logarithmic contribution 
to the DPS cross section associated with diagrams such as those in figure \ref{fig:genericcrosstalk} in which the 
two nonperturbatively generated ladders talk to one another by exchanging partons, provided that the crosstalk 
occurs at a lower scale than the scale of the $1 \to 2$ ladder branching. There are two types of crosstalk that
are possible, which are illustrated in the simple diagrams in figure \ref{fig:recombDPS}(a) and (b) - we'll call
these off-diagonal real emission and virtual exchange processes respectively. As in the previous section, we'll 
demonstrate that there is a leading logarithmic contribution from diagrams such as figure \ref{fig:genericcrosstalk} by
examining one of the simplest possible diagrams of the appropriate type -- namely, that of figure \ref{fig:recombDPS}(a). 
We will find that there is a large DGLAP logarithm associated with both the $1 \to 2$ splitting and the off-diagonal 
real emission (`crosstalk') processes in the figure, and that this is associated exclusively with the region of 
integration in which the partonic products of the off-diagonal real emission have much smaller transverse momentum 
than the products of the $1 \to 2$ splitting (and all of these transverse momenta are $\gg \Lambda^2$ but $\ll Q^2$).

In our calculation, we'll ignore considerations of colour for simplicity, just as we did in section \ref{sec:2v1calc}.
However, the colour structure of crosstalk processes is quite nontrivial, and is important when considering the size
of such contributions to cross sections. The colour structure of crosstalk processes has been considered previously
in the context of twist-4 contributions to DIS in \cite{Levin:1992mu, Bartels:1992ym, Bartels:1993ih, Bartels:1993ke}, 
and in the context of DPS in \cite{Bartels:2011qi, Manohar:2012jr, Manohar:2012pe}. We will make some comments with
regards to the colour structure of the crosstalk processes at the end of this section.

As in section \ref{sec:2v1calc}, we work in a frame in which $A = A^+p, B=B^-n$. The cross section expression 
associated with figure \ref{fig:recombDPS}(a) is:
\begin{align} \label{XtalkXsec}
\sigma_{XT}(s) = \dfrac{1}{2(2\pi)^{10}s} \int & d^4 \bar{A} d^4 r d^4 J_1 d^4 J_2 \delta(J_1^2 - Q^2) \delta(J_2^2 - Q^2)
\delta(r^2) 
\\ \nonumber
& \delta^{(4)}(A+B-\bar{A}-r-J_1-J_2) \mathcal{M}_L^{\lambda_1;\mu_1\mu_2\mu_3\chi}(A,B;J_1,J_2,r,\bar{A})
\\ \nonumber
& \mathcal{M}_R^{\lambda_1;\mu_1\mu_2\mu_3\chi}(A,B;J_1,J_2,r,\bar{A})^*
\end{align}
where:
\begin{align}
\mathcal{M}_L^{\lambda;\mu_1\mu_2\mu_3\chi}&(A,B;J_1,J_2,r,\bar{A}) 
\\ \nonumber
&= i^9g_s^2(eQ_q)^2 \int \dfrac{d^4 a_1}{(2\pi)^4} 
\dfrac{\mathrm{Tr}\left[\slashed{\epsilon}_{\lambda}(B) \slashed{b}_1 \slashed{\epsilon}^*_{\mu_1} (J_1) \slashed{a}_1 \varphi_p^\chi(a_1,a_2,\bar{A}) \slashed{a}'_2 \slashed{\epsilon}^*_{\mu_3} (r) \slashed{a}_2 \slashed{\epsilon}^*_{\mu_2}(J_2) \slashed{b}_2 \right]}{D(a_1)D(b_1)D(a'_2)D(b_2)D(a_2)}
\\
\mathcal{M}_R^{\lambda_1;\mu_1\mu_2\mu_3\chi}&(A,B;J_1,J_2,r,\bar{A}) 
\\ \nonumber
&= i^9g_s^2(eQ_q)^2 \int \dfrac{d^4 \tilde{a}_1}{(2\pi)^4} 
\dfrac{\mathrm{Tr}\left[ \slashed{\epsilon}_{\lambda_1}(B) \slashed{\tilde{b}}_1 \slashed{\epsilon}^*_{\mu_1} (J_1) \slashed{\tilde{a}}_1 \slashed{\epsilon}^*_{\mu_3} (r) \slashed{\tilde{a}}'_1 \varphi_p^\chi(\tilde{a}_1,\tilde{a}_2,\bar{A}) \slashed{\tilde{a}}_2 \slashed{\epsilon}^*_{\mu_2}(J_2) \slashed{\tilde{b}}_2  \right]}
{D(\tilde{a}_1)D(\tilde{b}_1)D(\tilde{a}'_1)D(\tilde{b}_2)D(\tilde{a}_2)}
\end{align}

Following a procedure that is similar to that leading to equation \eqref{2v1ME4}, and is 
valid in the region of transverse momentum integration in which $\vect{J}_1^2, \vect{J}_2^2,
\vect{r}^2 \gg \Lambda^2$ (or equivalently $\vect{a}_2^2, \vect{b}_1^2, \vect{b}_2^2 \gg
 \Lambda^2$), we can write down the following approximate expression for $\mathcal{M}_L$:
\begin{align} 
\mathcal{M}_L^{\lambda;\mu_1\mu_2\mu_3\chi}&(A,B;J_1,J_2,r,\bar{A})
 \\ \nonumber 
\simeq & \sum_{s_1s'_2} \int \dfrac{d a_1^+}{2\pi} \left[ \int d^2\vect{a_1}/(2\pi)^2\psi_p^{s_1s'_2\chi}(a_1^+,a_2^{\prime+},\vect{a_1},\vect{a'_2},\bar{A}^-) \right]
\\ \nonumber 
& \times \dfrac{\mathcal{T}_L^{s'_2s_1\lambda;\mu_1\mu_2\mu_3}(a'_2a_1B \to J_1J_2r) |_{a_1^-= 0,\vect{a_1}=0}}{[2(J_1^+ -a_1^+)J_1^- - \vect{J}_1^2 + i\epsilon]
[2(a_1^+ - J_1^+)(B^- - J_1^-) - \vect{J}_1^2 + i\epsilon]}
\\ \nonumber
& \times \dfrac{1}{[2(J_1^++J_2^+-a_1^+)(-r^-) - \vect{r}^2 +i\epsilon]}
\end{align}
where $\mathcal{T}_L(a_2a_1B \to J_1J_2r)$ includes all of the numerator structure of the 
$\mathcal{M}_L(a_2a_1B \to J_1J_2r)$ matrix element.

Performing the $a^+_1$ integral using contour methods, and making use of the fact that the
overall integrand is strongly peaked near $a^+_1 = J^+_1$ whilst the numerator factor 
$\mathcal{T}_L$ is a relatively smooth function in this region, we obtain:
\begin{align} \label{MLfinal}
\mathcal{M}_L^{\lambda;\mu_1\mu_2\mu_3\chi}&(A,B;J_1,J_2,r,\bar{A})
 \\ \nonumber
\simeq & \sum_{s_1s'_2} \dfrac{-ir^+}{2} \left[ \int d^2\vect{a_1}/(2\pi)^2\psi_p^{s_1s'_2\chi}(J_1^+,a_2^{\prime+},\vect{a_1},\vect{a'_2},\bar{A}^-) \right]
\\ \nonumber
& \times \dfrac{\mathcal{T}_L^{s'_2s_1\lambda;\mu_1\mu_2\mu_3}(a'_2a_1B \to J_1J_2r) |_{a_1^+ = J_1^+, a_1^-= 0,\vect{a_1}=0}}
{\vect{J}_1^2\vect{r}^2[J_2^+ + \tfrac{\vect{J}_1^2}{2J_1^-}+r^+]}
\end{align}

In the region of integration in which $\vect{J}_1^2 \ll Q^2$, we can drop the second term 
in the denominator factor $[J_2^+ + \tfrac{\vect{J}_1^2}{2J_1^-}+r^+]$. Also, when 
$\vect{J}_1^2, \vect{J}_2^2, \vect{r}^2 \ll Q^2$, we can approximately decompose 
$\mathcal{T}_L$ as follows:
\begin{align} \label{TLfinal}
\mathcal{T}_L&^{s'_2s_1\lambda;\mu_1\mu_2\mu_3}(a'_2a_1B \to J_1J_2r) |_{a_1^-= 0,\vect{a_1}=0} \simeq 
\mathcal{M}^{s_1t_1;\mu_1}_{q\bar{q} \to \gamma^*}(J_1^+p,J_1^-n \to J_1)
\\ \nonumber
& \times \mathcal{M}^{s_2t_2;\mu_2}_{\bar{q}q \to \gamma^*}(J_2^+p,J_2^-n \to J_2)
\mathcal{M}^{s'_2 \to s_2\mu_3}_{\bar{q} \to \bar{q}g}\left(a_2^{\prime+}p, a_2^+p-\vect{r}, (a_2^{\prime+}-a_2^+)p+\vect{r}\right)
\\ \nonumber
& \times \mathcal{M}^{\lambda \to t_1t_2}_{g \to \bar{q}q}(B;J_1^-n+\vect{J}_1,J_2^-n-\vect{J}_1) 
\end{align}

Performing a similar sequence of operations for $\mathcal{M}_R$, we obtain the expression
below that is valid for $\Lambda^2 \ll \vect{\tilde{a}}_1^2, \vect{\tilde{b}}_1^2, \vect{\tilde{b}}_2^2 \ll Q^2$, or equivalently $\Lambda^2 \ll \vect{J}_1^2, \vect{J}_2^2, \vect{r}^2 \ll Q^2$:
\begin{align} \label{MRfinal}
\mathcal{M}_R^{\lambda;\mu_1\mu_2\mu_3\chi}&(A,B;J_1,J_2,r,\bar{A})
 \\ \nonumber
\simeq & \sum_{s_i} \dfrac{-ir^+}{2} \left[ \int d^2\vect{\tilde{a}'_1}/(2\pi)^2\psi_p^{\tilde{s}'_1\tilde{s}_2\chi}(\tilde{a}_1^{\prime+},J_2^+,
\vect{\tilde{a}'_1}, \vect{\tilde{a}_2},\bar{A}^-) \right]
\\ \nonumber
& \times \dfrac{\mathcal{M}^{\tilde{s}'_1 \to \tilde{s}_1\mu_3}_{q \to qg}\left(\tilde{a}_1^{\prime+}p, a_1^+p-\vect{r}, (\tilde{a}_1^{\prime+}-a_1^+)p+\vect{r}\right)}
{\vect{J}_2^2\vect{r}^2[J_1^+ + \tfrac{\vect{J}_2^2}{2J_2^-}+r^+]}
\\ \nonumber
& \times \mathcal{M}^{\tilde{s}_1\tilde{t}_1;\mu_1}_{q\bar{q} \to \gamma^*}(J_1^+p,J_1^-n \to J_1)
\mathcal{M}^{\tilde{s}_2\tilde{t}_2;\mu_2}_{\bar{q}q \to \gamma^*}(J_2^+p,J_2^-n \to J_2)
\\ \nonumber
& \times \mathcal{M}^{\lambda \to \tilde{t}_1\tilde{t}_2}_{g \to \bar{q}q}(B;J_1^-n-\vect{J}_2,J_2^-n+\vect{J}_2)
\end{align}

Given that the transverse momenta of the partons emerging from the $g \to q\bar{q}$ branching
process are different on the left and right hand sides of the cut in figure \ref{fig:recombDPS}(a) ($\pm \vect{J_1}$
and $\mp \vect{J_2}$ respectively), we will require a generalised version of the relation
\eqref{1->2sptfn}, which reads:
\begin{align} \label{1->2sptfngen}
\dfrac{J_1^-J_2^-}{(J_1^-+J_2^-)^2} &\mathcal{M}^{\lambda \to t_1t_2}_{g \to \bar{q}q}(B;J_1^-n+\vect{J}_1,J_2^-n-\vect{J}_1)
\mathcal{M}^{*\lambda \to \tilde{t}_1\tilde{t}_2}_{g \to \bar{q}q}(B;J_1^-n-\vect{J}_2,J_2^-n+\vect{J}_2)
\\ \nonumber
&= -4g_s^2P_{g \to q\bar{q}}^{\lambda \to t_2t_1,\tilde{t}_2\tilde{t}_1}\left(\dfrac{J_2^-}{J_1^-+J_2^-}\right) \vect{\epsilon}_{\lambda} \cdot \vect{J}_1 
\vect{\epsilon}^*_{\lambda} \cdot \vect{J}_2 
\end{align}

Note that in the off-diagonal emission process, the partons emitting the gluon in the 
amplitude and conjugate do not in general have the same plus momentum (and indeed are not of 
the same type). This means that the product of $\mathcal{M}_{\bar{q} \to \bar{q}g}$ and
$\mathcal{M}^*_{q \to q g}$ from the left and right hand sides of the diagram does not give rise
to a conventional splitting function multiplied by the appropriate transverse momentum squared,
as occurred in \eqref{1->2sptfn}. Instead, one obtains:
\begin{align} \label{qqsptfngen}
\dfrac{r^+A^+}{\tilde{a}^{\prime+}_1a_2^{\prime+}}\sqrt{\dfrac{a_1^+a_2^+}{\tilde{a}^{\prime+}_1a_2^{\prime+}}}&\mathcal{M}^{s'_2 \to s_2\mu_3}_{\bar{q} \to \bar{q}g}\left(a_2^{\prime+}p, a_2^+p-\vect{r}, (a_2^{\prime+}-a_2^+)p+\vect{r}\right)
\\ \nonumber
\times &\mathcal{M}^{*\tilde{s}'_1 \to \tilde{s}_1\mu_3}_{q \to q g}\left(\tilde{a}_1^{\prime+}p, a_1^+p-\vect{r}, (\tilde{a}_1^{\prime+}-a_1^+)p+\vect{r}\right) 
\\ \nonumber
& \equiv 2g_s^2 
V_{I, q \to q}^{\tilde{s}'_1s'_2 \to \tilde{s}_1s_2;\mu_3} \left( \dfrac{a_1^+}{A^+},
\dfrac{\tilde{a}_1^{\prime+}}{A^+} , \dfrac{a_2^{\prime+}}{A^+} 
\right) \vect{r}^2
\end{align}
where $V_{I, q \to q}^{\tilde{s}'_1s'_2 \to \tilde{s}_1s_2;\mu_3} \left( \tfrac{a_1^+}{A^+},
\tfrac{\tilde{a}_1^{\prime+}}{A^+} , \tfrac{a_2^{\prime+}}{A^+} \right)$ represents some 
kind of generalised splitting function, that satisfies the following relation:
\begin{align}
V_{I, q \to q}^{\tilde{s}'_1s'_2 \to \tilde{s}_1s_2;\mu_3} \left( \dfrac{a^+}{A^+},
 \dfrac{a^{\prime+}}{A^+} , \dfrac{a^{\prime+}}{A^+} \right)
= \dfrac{A^+}{a^{\prime+}} P_{qq}^{\tilde{s}'_1s'_2 \to \tilde{s}_1s_2;\mu_3}\left(\dfrac{a^+}{a^{\prime+}}\right)
\end{align}

Furthermore, since the partons emerging from the hadronic blob in figure \ref{fig:recombDPS}(a) do not in
general carry the same momentum on the left and right hand sides of the diagram, the 
process in figure \ref{fig:recombDPS}(a) probes a two-parton PDF that is not diagonal in $x$. It is defined
according to:
\begin{align} \label{tw4distro}
 \Gamma_{p;q\bar{q}}^{s_1s_2,\tilde{s}_1\tilde{s}_2} \biggl( 
 \dfrac{a_1^+}{A^+},\dfrac{a_2^{\prime+}}{A^+}, &
 \dfrac{\tilde{a}_1^{\prime+}}{A^+} \biggr) \equiv \dfrac{2}{(2\pi)^9}
\sum_\chi\int  d\bar{A}^- d^2\vect{\bar{A}}d^2\vect{a_1}d^2\vect{\tilde{a}'_1}  \sqrt{a_1^+a_2^{\prime+}\tilde{a}_1^{\prime+}\tilde{a}_2^{+}}A^+
\\ \nonumber
& \times \psi_{p;q\bar{q}}^{s_1s_2\chi}(a_1^+,a_2^{\prime+},\vect{a}_1,\vect{a}'_2,\bar{A}^-,\vect{\bar{A}})
\psi_{p;q\bar{q}}^{*s'_1s'_2\chi}(\tilde{a}_1^{\prime+},\tilde{a}_2^{+},\vect{\tilde{a}'_1},\vect{\tilde{a}_2},\bar{A}^-,\vect{\bar{A}})
\end{align}

Note that this distribution is somewhat similar to the four-quark matrix element that is 
probed in the twist-four contribution to Drell-Yan, and that is defined in \cite{Ellis:1982wd, 
Ellis:1982cd, Qiu:1988dn, Qiu:1990xxa}. Here, however, we do not absorb two powers of the 
strong coupling constant $g_s$ into the four quark matrix element, as is done (and makes sense)
in the context of the twist-four contribution to Drell-Yan.

Inserting \eqref{MLfinal}, \eqref{TLfinal}, and \eqref{MRfinal} into \eqref{XtalkXsec}, and
making use of \eqref{1->2sptfngen}, \eqref{qqsptfngen} and \eqref{tw4distro}, we find that the
contribution to $\sigma_{XT}$ coming from the region of transverse momentum integration with
$\Lambda^2 \ll \vect{r}^2, \vect{J}_1^2, \vect{J}_2^2 \ll Q^2$ is:
\begin{align} \label{XtalkXsec2}
\sigma_{XT}(s) =& \sum_{s_i \tilde{s}_i t_i \tilde{t}_i \tilde{s}'_1 s'_2}\int dx_1 dx_2 dy_1 dy_2 
\hat{\sigma}^{s_1,t_1;\tilde{s}_1,\tilde{t}_1;\mu_1}_{\bar{q}q \to \gamma*}(\hat{s} = x_1y_1s)
\hat{\sigma}^{s_2,t_2;\tilde{s}_2,\tilde{t}_2;\mu_2}_{q\bar{q} \to \gamma*}(\hat{s} = x_2y_2s)
\\  \nonumber
& \left[\dfrac{\alpha_s}{2\pi} \int_{x_1}^{1-x_2} d\tilde{x}'_1 V_{I, q \to q}^{\tilde{s}'_1s'_2 \to \tilde{s}_1s_2;\mu_3}(x_1,\tilde{x}'_1,x'_2) \Gamma_{p;q\bar{q}}^{s_1s'_2,\tilde{s}'_1\tilde{s}_2}\left(x_1,x'_2,\tilde{x}'_1\right) \right]
\\  \nonumber
& \left[\dfrac{\alpha_s}{2\pi}P_{g \to q\bar{q}}^{\lambda \to t_2t_1,\tilde{t}_2\tilde{t}_1}\left(y_2\right)
\delta(1-y_1-y_2) \right] \int d\vect{J_1}^2 d\vect{r}^2 \dfrac{2\vect{\epsilon}_{\lambda} \cdot \vect{J}_1 
\vect{\epsilon}^*_{\lambda} \cdot (\vect{J}_1 + \vect{r}) }{\vect{r}^2\vect{J}_1^2 (\vect{J}_1 + \vect{r})^2}
\end{align}

In the region of transverse momentum integration in which $\vect{r}^2 \ll \vect{J}_1^2, \vect{J}_2^2$,
the transverse momentum integrand simplifies as below, and we obtain two large DGLAP logarithms from 
this region:
\begin{align}
\int d\vect{J_1}^2 d\vect{r}^2 \dfrac{2\vect{\epsilon}_{\lambda} \cdot \vect{J}_1 \vect{\epsilon}^*_{\lambda} 
\cdot (\vect{J}_1 + \vect{r}) }{\vect{r}^2\vect{J}_1^2 (\vect{J}_1 + \vect{r})^2}
\xrightarrow[\vect{J_1}^2 \gg \vect{r}^2]{}
\int_{\Lambda^2}^{Q^2} \dfrac{d\vect{J_1}^2}{\vect{J_1}^2} 
\int_{\Lambda^2}^{\vect{J_1}^2} \dfrac{d\vect{r}^2}{\vect{r}^2} = \log^2\left( \dfrac{Q^2}{\Lambda^2} \right)
\end{align}

Two large DGLAP logarithms implies a leading logarithmic contribution, since there are two powers of
$\alpha_s$ in \eqref{XtalkXsec2}. Thus, there is a leading logarithmic contribution to the DPS cross
section coming from the region of figure \ref{fig:recombDPS}(a) in which $\vect{r}^2 \ll \vect{J}_1^2$ (i.e. in which the
scale of the off-diagonal real emission process is strictly smaller than the scale of the $1 \to 2$
branching process). It is only this region of transverse momentum integration that gives rise to a 
leading double logarithm -- other regions only give rise to either a single logarithm, or no logarithm 
at all. The single DGLAP logarithm is essentially associated with a logarithmic integral over $\vect{r}$ 
only, and this should be absorbed into the four-quark matrix element in the `conventional' twist-4 
contribution to double Drell-Yan.

In \eqref{XtalkXsec2} we have not explicitly written the scales associated with the two factors of 
$\alpha_s$. It should be clear that in the integration region that gives rise
to the leading logarithm, the appropriate scale for the first $\alpha_s$ factor should be $\vect{r}^2$, 
whilst that for the second factor should be $\vect{J}_1^2$ (with $\vect{r}^2 \ll \vect{J}_1^2$ in the 
integration region of interest). This is because the first coupling constant is associated with the
crosstalk interaction (that gives rise to transverse parton momenta of order $\vect{r}^2$), whilst
the second is associated with the $1 \to 2$ splitting (that gives rise to transverse parton momenta
of order $\vect{J}_1^2$).

Aside from the process in figure \ref{fig:recombDPS}(a) involving an off-diagonal real emission, the
process in figure \ref{fig:recombDPS}(b) involving a virtual exchange also gives rise to a leading 
double logarithm, provided once again that the virtual exchange process occurs at a lower scale
than the $1 \to 2$ branching. This is straightforward to show using a procedure similar to the one 
we have used above. Generalising these results, we find that in the most general 2v1 DPS diagram, 
all possible types of parton exchange are allowed inside the two ladders emerging from one of the 
protons at leading logarithmic order, provided that they occur at a lower scale than the $1 \to 2$ 
ladder branching occurring in the other proton. Schematically, the LO (diagonal unpolarised) cross 
section expression for the 2v1 contribution to DPS is:
\begin{align} \label{2v1Xtalk}
\sigma^{D,2v1}_{(A,B)}(s) =& 2 \times \dfrac{m}{2} \sum_{l i_i j_i i'_i j'_i} \int_{\Lambda^2}^{Q^2} dk^2 \dfrac{\alpha_s(k^2)}{2\pi k^2} \int dx_1 dx_2 dy_1 dy_2 
\dfrac{dx'_1}{x'_1} \dfrac{dx'_2}{x'_2} \dfrac{dy'_1}{y'_1} \dfrac{dy'_2}{y'_2}
\\ \nonumber
& \times \hat{\sigma}_{i_1j_1 \to A}(\hat{s} = x_1y_1s)
\hat{\sigma}_{i_2j_2 \to B}(\hat{s} = x_2y_2s)
\\ \nonumber
& \times \dfrac{D^{l}_p(y'_1+y'_2,k^2)}{y'_1+y'_2} P_{l \to j_1'j_2'}\left(\dfrac{y'_1}{y'_1+y'_2}\right)
D_{j_1'}^{j_1}\left(\dfrac{y_1}{y'_1};k^2,Q^2\right) D_{j_2'}^{j_2}\left(\dfrac{y_2}{y'_2};k^2,Q^2\right)
\\ \nonumber
& \times D_{i_1'}^{i_1}\left(\dfrac{x_1}{x'_1};k^2,Q^2\right) D_{i_2'}^{i_2}\left(\dfrac{x_2}{x'_2};k^2,Q^2\right)
\Gamma^{i_1'i_2'}_p(x'_1, x'_2;x'_1, k^2)
\end{align}
$\Gamma^{i_1'i_2'}_p(x_1, x_2;\tilde{x}_1, \mu^2)$ is a four-parton matrix element whose evolution
involves all possible exchanges between these partons in an axial gauge -- i.e. the two types of 
real emission plus virtual exchange and self energy corrections\footnote{In a covariant gauge, such 
as Feynman gauge, there are further diagrams that contribute to the evolution due to the presence of a 
nontrivial Wilson line in the definition of the operator. These diagrams involve gluon connections to 
the Wilson line.}. Taking Mellin moments of this function gives rise to a matrix element of one of the 
so-called `quasipartonic operators' whose evolution is discussed in \cite{Bukhvostov:1985rn}. Note that
taking Mellin moments of a 2pGPD normally does not give rise to the expectation value of a quasipartonic
operator, due to the finite $\vect{b}$.

It is straightforward to show that if the crosstalk interactions are omitted, then equation \eqref{2v1Xtalk}
reduces to \eqref{1v2XSec}. With crosstalk interactions absent, $\Gamma^{i_1'i_2'}_p(x'_1, x'_2;x_1', 
k^2)$ in \eqref{2v1Xtalk} is built up from two independent ladders:
\begin{equation}
\Gamma^{i_1'i_2'}_p(x'_1, x'_2;x_1', k^2) = \Gamma^{i_1'i_2'}_{p,indep}(x'_1, x'_2, \vect{b} = \vect{0}; k^2)
\end{equation}
where $\Gamma^{i_1'i_2'}_{p,indep}$ is given by \eqref{indepbranchGPD}. Substituting this expression for 
$\Gamma^{i_1'i_2'}_p(x'_1, x'_2;x_1', k^2)$ into \eqref{2v1Xtalk}, and making use of the relation:
\begin{equation}
\sum_{j} \int_x^1 \dfrac{dx'}{x'} D_{i}^{j} \left(x';\Lambda^2,k^2\right) D_{j}^{k} \left(\dfrac{x}{x'};k^2,Q^2\right)
= D_{i}^{k} \left(x;\Lambda^2,Q^2\right)
\end{equation}
we observe that the expression for $\sigma^{D,2v1}_{(A,B)}$ becomes equal to \eqref{1v2XSec}.

At the next-to-leading logarithmic (or NLO) level, one would need to append an extra term to 
\eqref{2v1Xtalk} that is of the following form:
\begin{align}
\int dx_1 dx_2 d\tilde{x}_1 dy D^{k}_p (y,Q^2) \Gamma^{ij}_p(x_1, x_2;\tilde{x}_1, Q^2) \hat{\sigma}_{ijk \to AB} 
(x_1,x_2,\tilde{x}_1,y)
\end{align}

This is essentially the `conventional' twist-4 contribution to the $pp \to AB + X$ production cross 
section. At the level of total cross sections, the DPS contribution to the production of $AB$ cannot
be distinguished from the conventional twist-4 contribution, and the two should really just be considered 
together as components of the $\mathcal{O}(\Lambda^2/Q^2)$ correction to the $pp \to AB + X$ cross section.

Let us now discuss the issue of colour in the evolution of the four-parton (twist-4) matrix element 
$\Gamma^{i_1'i_2'}_p(x_1, x_2;\tilde{x}_1, \mu^2)$. We recall that, for the 2pGPD with finite $\vect{b}$,
every distribution that does not have the partons with the same light-cone momentum fractions on either 
side of the cut paired up into colour singlets is suppressed by a Sudakov factor -- see \cite{Mekhfi:1988kj, 
Diehl:2011tt, Diehl:2011yj, Manohar:2012jr}. This factor arises in axial gauge because there is an incomplete 
cancellation of the soft gluon region between (diagonal) real emission diagrams and virtual self-energy 
corrections in the colour interference/correlation distributions \cite{Mekhfi:1988kj}. In physical 
terms, it occurs because such distributions involve a movement of colour by the large transverse distance 
$\vect{b}$ in the hadron \cite{Manohar:2012jr}. 

In the twist-4 matrix element $\Gamma^{i_1'i_2'}_p(x_1, x_2;\tilde{x}_1, \mu^2)$ there is no such Sudakov
suppression of colour interference/correlation distributions. The extra diagrams that are allowed in the
evolution of this distribution (i.e. the off-diagonal emission and virtual exchange diagrams in axial gauge) 
provide extra soft-gluon divergences that cancel any remaining divergence from adding the diagonal real emission 
and virtual self-energy diagrams together. The soft divergence in both real emission diagrams (diagonal and off-diagonal)
is positive, whilst that in both virtual diagrams (self-energy and exchange) is negative, and in the sum the
positive and negative contributions always cancel each other out. We can see why this cancellation occurs 
physically as follows. In the operator definition of the twist-4 matrix element, the four operators corresponding 
to the partons all lie on the same lightlike line, with no transverse separation between any of them. Note that this
does not exactly correspond to the physical situation that we have in the 2v1 graphs -- in these, the transverse 
separation of the partons in the nonperturbatively generated pair must be equal to that of the partons emerging 
from the $1 \to 2$ splitting in the other proton, which is of order $1/k$ (with $k^2$ equal to the scale of the 
$1 \to 2$ splitting). However, for the purposes of obtaining $\Gamma^{i_1'i_2'}_p(x'_1, x'_2;x'_1, k^2)$ in 
\eqref{2v1Xtalk} by solving the evolution equation at scales $\mu^2 < k^2$, this separation is not resolvable
and effectively can be taken as zero. The fact that the four operators/partons in $\Gamma^{i_1'i_2'}_p(x_1, 
x_2;\tilde{x}_1, \mu^2)$ are on top of (or at least very close to) one another in transverse space means that
soft longwave gluons can only resolve the total colour of all of them. But the summed colour of the four partons
must be zero, since the proton is a colour singlet object -- therefore the effects of soft gluons must cancel, as
is indeed observed in practical calculations. The cancellation of soft gluon divergences in the twist-4 matrix 
elements has been discussed before, in \cite{Gribov:1983joadd, Gribov:1988joadd, Levin:1992mu} (for example).

\begin{figure}
\centering
\includegraphics[scale=0.65]{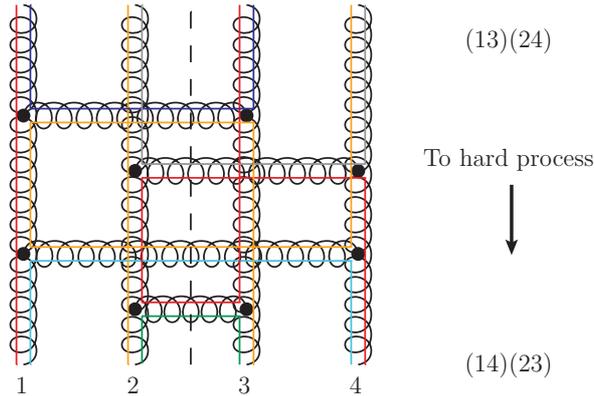}
\caption{A process that can bring about a colour recombination in the four gluon state. On the diagram we have
indicated the colour flow in the large $N_C$ limit.}
\label{fig:colourrecomb}
\end{figure}

It is important to point out that we are not claiming here that colour interference contributions to the 2v1
DPS cross section are free from Sudakov suppression. Rather, we are stating the well-known theoretical fact 
that there can be no Sudakov suppression in any twist-4 matrix element. In the 2v1 DPS cross section there 
can be a Sudakov suppression if the partons coming from either proton are not colour matched between amplitude
and conjugate at scales $\mu^2 > k^2$ at which the parton transverse separation $\sim 1/k^2$ is resolvable. 
On the two initial-state ladder side, and at scales $\mu^2 < k^2$ at which the parton transverse separation is
not resolvable, the way in which colour is distributed between the parton legs in amplitude and conjugate can
change with no accompanying Sudakov suppression.

Let us now consider the region of small $x$ in the four-parton matrix element (which one might believe to be
the most relevant region for DPS at the LHC -- but see later). It is well known that in this region the gluons 
dominate, so we will only consider these partons in what follows. We have seen that the colour 
correlated/interference twist-4 distributions are not Sudakov suppressed -- however, in the low $x$ region, 
the distributions in which two pairs of gluons are in colour singlet configurations tend to win out. This is 
because the colour factors in the anomalous dimensions for these distributions are larger (see section 3.2 of 
\cite{Levin:1992mu} or section 5.1.3 of \cite{Diehl:2011yj}). Bear in mind that in figure \ref{fig:genericcrosstalk} 
at scale $k^2$ the nonperturbatively generated partons with identical $x$ fractions must be in a colour singlet 
state if one wants to avoid any Sudakov suppression. By combining two off-diagonal real emission processes put 
together with two diagonal real emission processes as in figure \ref{fig:colourrecomb}, it is possible to 
achieve a `colour recombination' on the two ladder side at scales lower than $k^2$, and alter the way in 
which the parton legs are grouped into two sets of colour singlets. For example, in figure \ref{fig:colourrecomb} 
the grouping is changed from (14)(23) afterwards to (13)(24) before, using the leg labelling conventions from 
the figure. Such a colour recombination is not disfavoured from the point of view of evolution before and 
after the process -- however, it is itself suppressed by a colour factor equal to $1/(N_C^2 - 1)$ 
\cite{Levin:1992mu, Bartels:1992ym, Bartels:1993ih, Bartels:1993ke, Bartels:2011qi}. This colour factor 
suppression is associated with the fact that the recombination process is non-planar. 

From the point of view of low $x$ physics, there is an important distinction between the two crosstalk 
processes that we have discussed in this section -- i.e. the off-diagonal real emission and virtual exchange
processes. The off-diagonal real emission process can significantly reduce the magnitudes of the lightcone
momentum fractions of the two active parton legs involved in the process, since it is a real emission process.
On the other hand, the same is not true for the virtual exchange process. Here, the sum of the lightcone 
momentum fractions of the two parton legs involved must be conserved, and since the two legs are forced to
have positive lightcone momentum fractions by the kinematics of the process, the magnitudes of both $x$s 
cannot simultaneously decrease -- one must increase to compensate the decrease of the other. This means that,
taking all partons involved to be gluons as is appropriate at low $x$, the virtual exchange splitting function 
is not enhanced at small $x$ in the same way that the off-diagonal exchange (and indeed diagonal exchange) 
splitting functions are. In particular, the virtual exchange diagrams do not contribute at double leading 
logarithmic order to the evolution of the four-gluon matrix element. This result has been known for some time 
-- see \cite{McCoy:1974qb, McCoy:1974qc, Matinyan:1976aa, Matinyan:1976ab, Matinyan:1976ac, Levin:1992mu}. This 
is the reason why we drew the colour recombination process in figure \ref{fig:colourrecomb} using two off-diagonal 
real emission processes -- it would be also possible to engineer a colour recombination using two virtual exchange 
processes instead, but such a process would not be as strongly enhanced at low $x$.

We now discuss the important question of the numerical impact of the crosstalk interactions on the 2v1 DPS cross section. A complete investigation of
this issue at leading logarithmic order in QCD would require taking into consideration all partonic, spin and colour
channels, and the calculation and numerical implementation of all of the evolution kernels between these channels at
leading order. This is beyond the scope of this paper -- however, we can make a few important comments. 

One might expect the effects of the crosstalk interactions to be negligible based on the colour suppression of these
interactions in the twist-4 matrix element that we mentioned above. On the other hand, in \cite{Bartels:1993ke}, an
investigation into the size of crosstalk effects in the four-gluon matrix element was performed using the double 
leading logarithmic approximation (and in the context of shadowing corrections to DIS), and it was found
that the size of the effects is appreciable even for not too small $x$ and evolution lengths. It was shown that the
effects of the crosstalk interactions on the four-gluon matrix element could be approximately described by the following
`K-factor' (here we include the effects of the running coupling):
\begin{equation} \label{KfacBR}
K_{BR}(Y, \xi )= 1 + 2 \sqrt{\pi \delta} \left( \dfrac{4N_C}{\pi b}Y\ln\left(\dfrac{\xi-\xi_{\Lambda}}{-\xi_{\Lambda}}\right)\right)^{1/4}
\end{equation}
where $\delta \sim 1/N_C^4$, $b = (33-2N_f)/(12\pi)$, $Y = \ln(1/x)$, $\xi = \ln(Q^2/Q_0^2)$, $\xi_{\Lambda} = 
\ln(\Lambda_{QCD}^2/Q_0^2)$, $Q_0^2$ is the scale at which one begins including crosstalk effects, $Q^2$ is the 
final scale, and $x$ corresponds to the size of the $x$ values in the matrix element. If one plugs the sample 
values $x=0.001$, $Q = 5$ GeV, $Q_0 = 1$ GeV, $\Lambda_{QCD} = 0.359$ GeV, $N_f=3$ into this formula, one obtains
$K_{BR}=1.96$ - a significant enhancement despite the $1/(N_C^2-1)$ suppression of the recombination vertex. Thus,
we cannot simply ignore recombination processes in the 2v1 DPS cross section based only on their colour suppression.

An important point to emphasise in the context of numerical considerations is that the 2v1 contribution to DPS does not probe the twist-4 matrix 
elements at $\mu^2 = Q^2$ and $x \sim x_h$ (where $x_h$ represents the values of $x$ associated with the hard
scatterings). Rather, the twist-4 matrix elements are probed at $\mu^2 = k^2$ and $x = x_{sp}$, where $x_{sp}$
is the $x$ value appropriate to the $1 \to 2$ splitting. In general we will have $k^2 < Q^2$, $x_{sp} > x_h$, which
gives a smaller `evolution space' in $x$ and $\mu^2$ for the crosstalk effects, and a smaller enhancement.

Of course, there will not be a single value of $k^2$ and $x_{sp}$ that applies to all 2v1 graphs that have a 
particular hard scale $Q^2$ and particular values of $x$ at that hard scale -- instead we integrate over $k^2$ 
and $x_{sp}$, as one can see from looking at the equation \eqref{2v1Xtalk} ($x_{sp} = y_1' + y_2'$ in this 
equation). Two factors determine the average values of $Y_{split}$ and $\xi_{split}$ in a particular set of 
2v1 graphs -- the first is that we have two ladders rather than one after the split on the one ladder side
(which will tend to prefer smaller $Y_{split}$ and $\xi_{split}$), and the second is that we have no crosstalk
interactions after the split on the two ladder side (which will tend to prefer larger $Y_{split}$ and $\xi_{split}$).
One expects the first effect to be strongly dominant, in which case we can estimate $\langle Y_{split} \rangle$ and $\langle\xi_{split}\rangle$
for particular final $x$ and $Q^2$ values by considering just the one-ladder side. For example, we can estimate 
$\langle Y_{split} \rangle(x,Q^2)$ by weighting the integral \eqref{Dbreve} by $\ln(1/(y_1'+y_2'))$, setting $y_1 \simeq y_2 \simeq
x$, and then dividing by $\breve{D}_p(x,x;Q^2)$.

Using this method, we estimated the values of $\langle Y_{split} \rangle$ and $\langle\xi_{split}\rangle$ for various values of $x$ at the hard
interaction, and for $Q = 10$ GeV. In our approximate calculation we included only gluons, taking as input the
MSTW2008LO gluon input at $Q_0 = 1$ GeV (and only integrating $k^2$ down to $Q_0^2$ in \eqref{Dbreve} and the 
weighted integral). The number of flavours was held fixed, $N_f = 3$, we take $\Lambda_{QCD} = 0.359$ GeV, and \
we used the full $g \to g$ splitting functions.

The results of this calculation are given in table \ref{tab:spvals}. Rather than $\langle Y_{split} \rangle$ and 
$\langle \xi_{split} \rangle$, we have chosen to tabulate the values of $x$ and $\mu^2$ corresponding to these
values -- i.e. $\exp\left(-\langle Y_{split} \rangle \right)$ and $Q_0 \sqrt{\exp(\langle\xi_{split}\rangle)}$
respectively. We also tabulate the values of $K_{BR}$ corresponding to each set of $(\langle Y_{split} \rangle,
\langle \xi_{split} \rangle)$ values.

One notices immediately from the table that the $x$ values of splitting are much larger than that at the hard
process, and the $Q$ values of splitting tend to be rather close to $Q_0$ (as was also noticed in \cite{Ryskin:2012qx}). 
For $x \gtrsim 10^{-3}$, the $x$ values at the splitting begin to enter the region $\gtrsim 10^{-1}$ in which PDFs decrease with 
$\mu^2$ rather than increase (as the dominant process becomes splitting to smaller $x$ values rather than the
`feed' from larger $x$ values). For such $x$ values, the net effect of including crosstalk interactions should
either be small, or should actually be to reduce the 2v1 cross section (as the crosstalk interactions offer 
extra pathways by which the four-gluon matrix element can decrease rather than increase). Clearly, we cannot
place any trust at all in the $K_{BR}$ values quoted in table \ref{tab:spvals} for $x \gtrsim 10^{-3}$. 

For much smaller $x$ values (e.g. $10^{-6}$) the $x$ values at the splitting are smaller (although not by 
much), and one might anticipate an enhancement of the 2v1 cross section due to crosstalk interactions, 
although perhaps not as large as the formula \eqref{KfacBR} predicts. One must bear in mind, however, that for
such small $x$ values the DGLAP approach we have taken here may not be so well justified, and an alternative
approach based on the BFKL equation may be more appropriate.

\begin{figure}
\includegraphics[scale=0.65, trim = 0.8cm 0 0 0]{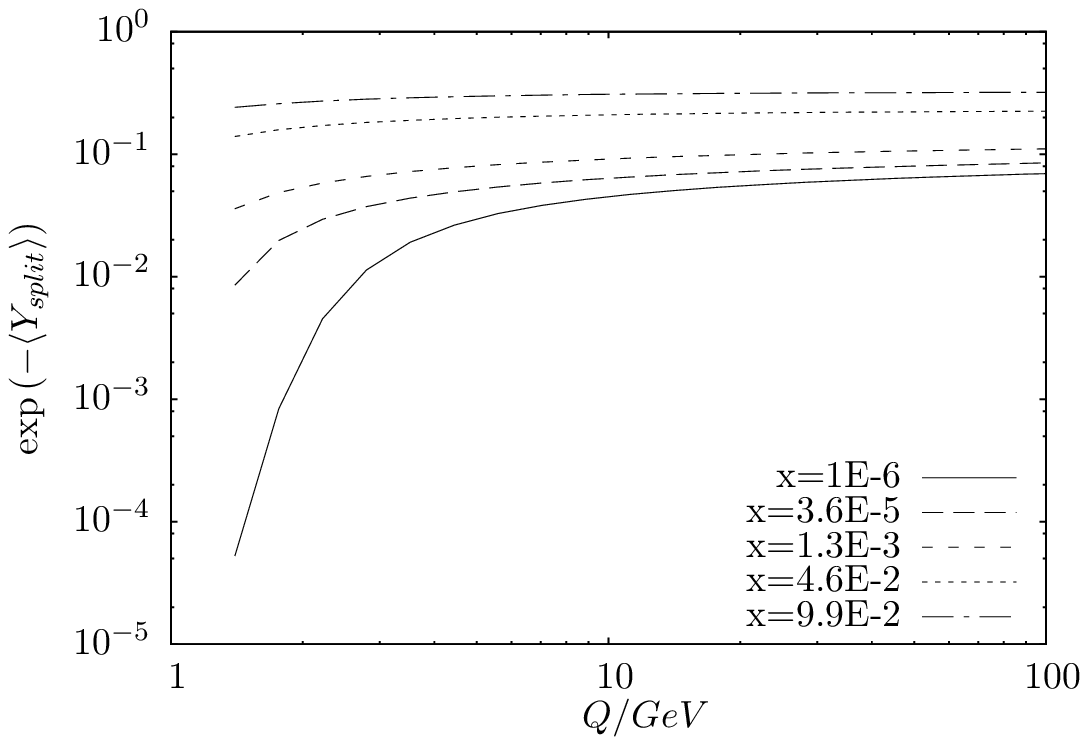}
\includegraphics[scale=0.65]{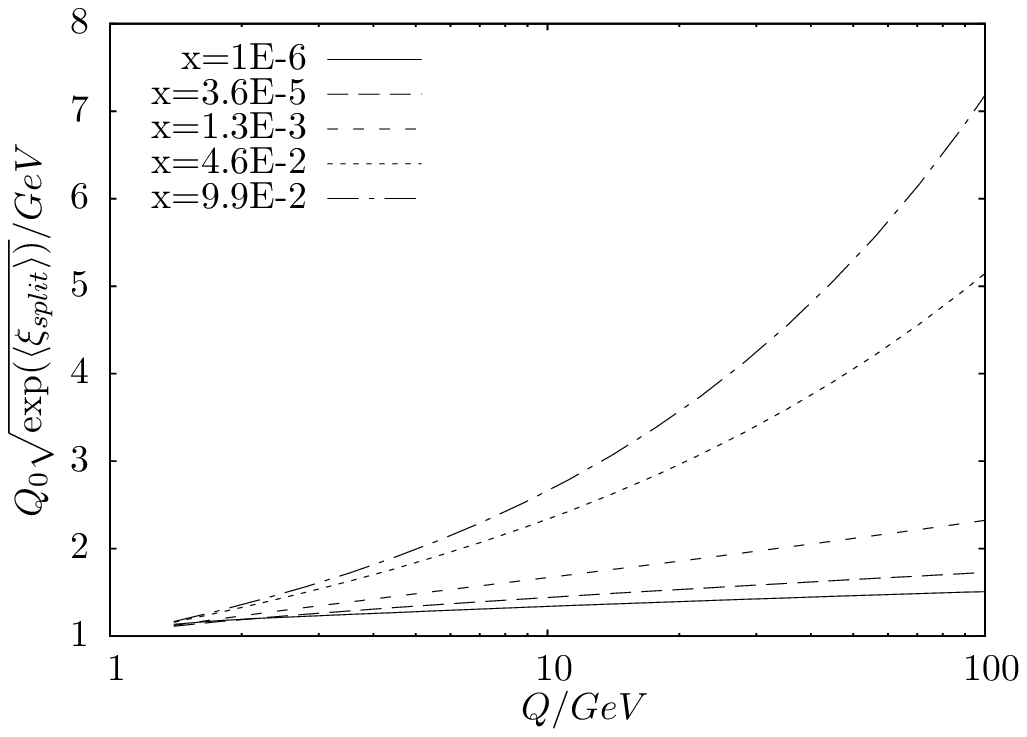}
\caption{Plots showing the $x$ and $\mu^2$ values corresponding to the average $Y$ and $\xi$ values of the $1 \to 2$
splitting, plotted against $Q^2$, and for various values of $x$ at the hard interaction.}
\label{fig:splitvals}
\end{figure}

Note that the results we obtained were for $Q = 10$ GeV. The $x$ and $\mu^2$ values corresponding to 
$\langle Y_{split} \rangle$ and $\langle\xi_{split}\rangle$ are plotted against $Q^2$ for various values of $x$
in figure \ref{fig:splitvals}. We notice that the typical $x$ and $\mu^2$ values of splitting increase with 
$Q^2$, although beyond $Q = 10$ GeV the typical $x$ values do not increase by much, and one would expect the 
conclusions we have found above to also hold for larger $Q$ values.

Thus, we have provided an indication that for not too small $x$ values ($ \gtrsim 10^{-3}$), crosstalk 
interactions should not significantly enhance the cross section, and indeed they most likely reduce it. Moderate
$x$ values ($10^{-3} - 10^{-2}$) correspond to $x_{sp}$ values $\simeq 0.1$ where PDFs do not change significantly
with scale, so one might expect the effect of crosstalk effects on the DPS cross section to be small at such $x$
values -- however, a more complete numerical investigation, which is beyond the scope of this paper, is required
to make a more definite statement.

\begin{table}
\centering
\begin{tabular}{|c|c|c|c|}
\hline
$x$ & $ \exp\left(-\langle Y_{split} \rangle \right) $ &  $Q_0 \sqrt{\exp(\langle\xi_{split}\rangle)}$ & $K_{BR}(\langle Y_{split} \rangle, \langle \xi_{split} \rangle )$ \\
\hline
1.0E-06 & 4.5E-02 & 1.34 & 1.56 \\
3.6E-05 & 6.4E-02 & 1.44 & 1.57 \\
1.3E-03 & 9.1E-02 & 1.67 & 1.59 \\
4.6E-02 & 2.1E-01 & 2.35 & 1.59 \\
9.9E-02 & 3.1E-01 & 2.67 & 1.56 \\
\hline
\end{tabular}
\caption{Average $Y$ and $\xi$ values of the $1 \to 2$ splitting for a `perturbatively generated' pair of partons with the
$x$ values in the table, and probed at $Q = 10$ GeV. Note that we have actually tabulated the $x$ and $\mu^2$ values 
corresponding to these average $Y$ and $\xi$ values. The values of $K_{BR}$ corresponding to the average $Y$ and $\xi$ 
values are tabulated, but for reasons given in the text one should not place much trust in these figures.}
\label{tab:spvals}
\end{table}

The statement that the `$\vect{b} = \vect{0}$' twist-4 distributions probed in the 2v1 contribution to DPS evolve 
differently from the 2pGPDs with finite $\vect{b}$ has been made recently in Appendix A of \cite{Manohar:2012pe}. 
However, in this paper it is claimed that only the evolution of the colour correlation/interference distributions 
changes at $\vect{b} = \vect{0}$ -- we contend that the evolution of the colour diagonal/singlet distribution 
is also affected in an important way. In equation (A1) of \cite{Manohar:2012pe}, an evolution equation for the 
colour octet twist-4 $q\bar{q}$ distribution diagonal in $x$ fractions is proposed. However, the equation they 
propose involves only similar distributions diagonal in $x$ fractions on the right hand side -- in fact the 
correct evolution equation should contain more general distributions nondiagonal in $x$ on the right hand side, 
since the crosstalk processes that are allowed for $\vect{b} = \vect{0}$ will necessarily disrupt a diagonal/symmetric 
pattern of $x$ values.

\section{The Total Cross Section for Double Parton Scattering}  \label{sec:XsecResults}

In a previous publication \cite{Gaunt:2011xd} (see also \cite{Gaunt:2012wv}), we examined another 
class of diagrams that can contribute to the DPS cross section -- the `1v1' graphs. For the 
arbitrary `1v1' graph in figure \ref{fig:2v21v2}(c) with a total of $n$ QCD branching vertices 
in the amplitude or conjugate, we discovered that there is no natural piece of the diagram that 
is proportional to $\log(Q^2/\Lambda^2)^n/R_p^2$ and is associated with the transverse momenta 
inside the loop being strongly ordered on either side of the diagram (in fact, most
of the contribution to the total cross section expression for the graph comes from the region
of integration in which the transverse momenta of particles inside the loop are of 
$\mathcal{O}(\sqrt{Q^2})$). Based on this finding, we suggested that 1v1 graphs should not
contribute to the pp DPS cross section.

Combining this suggestion with the findings of sections \ref{sec:2v1calc} and \ref{sec:crosstalk}, 
we obtain the following formula for the total DPS cross section:
\begin{align} \label{DPSmaster}
\sigma^D_{(A,B)}(s) = \sigma^{D,2v2}_{(A,B)}(s) + \sigma^{D,2v1}_{(A,B)}(s)
\end{align}
with $\sigma^{D,2v2}_{(A,B)}(s)$ and $\sigma^{D,2v1}_{(A,B)}(s)$ being given by the expressions \eqref{2v2XSec}
and \eqref{2v1Xtalk} respectively\footnote{Note that this is really our prediction for the unpolarised diagonal 
contribution to the total DPS cross section when the scales of the two hard interactions are the same, 
$Q_A^2 = Q_B^2 = Q^2$. To generalise this result to unequal scales, one needs to change $Q^2$ to $Q_A^2$
in all Green's functions in \eqref{2v1Xtalk} involving a `1' index, change $Q^2$ to $Q_B^2$ in all Green's 
functions in \eqref{2v1Xtalk} involving a `2' index, change the upper limit of the $k^2$ integration to
$\text{min}(Q_A^2, Q_B^2)$, and perform a similar operation for the `2v2' contribution. As mentioned previously,
the contributions associated with spin polarisation (either longitudinal or transverse) and flavour interference 
are expected to have a similar structure to \eqref{DPSmaster}, whilst the colour correlation/interference and
parton type interference contributions should be suppressed by Sudakov factors, as is discussed in 
\cite{Mekhfi:1988kj, Diehl:2011tt, Diehl:2011yj, Manohar:2012jr}.}. 

This formula agrees with the DPS cross section formulae proposed in \cite{Manohar:2012jr, Manohar:2012pe}
and \cite{Blok:2011bu}, apart from the fact that in neither of these papers are the crosstalk effects in 
the 2v1 graphs taken into account correctly (in \cite{Blok:2011bu} they are omitted, whilst in
\cite{Manohar:2012jr, Manohar:2012pe} they are included in an incorrect fashion). Ryskin and Snigirev
\cite{Ryskin:2011kk} include an extra `1v1' contribution in their proposed cross section formula, which 
they argue can be validly included in the DPS cross section at suitably low $x$ \cite{Ryskin:2012qx}, 
but which should not be included at moderate to large $x$ values.

We would like to point out at this stage that there are two features in the equation \eqref{DPSmaster}
that are potentially concerning, and that might indicate that modifications to it may be required in order 
to correctly describe the DPS cross section.

The first issue is that we were originally expecting to obtain an expression for the DPS cross section 
looking something like \eqref{DPSXsecgen}, with the 2pGPDs in these formulae each having an interpretation 
in terms of hadronic operator matrix elements. Our proposed expression \eqref{DPSmaster} deviates somewhat 
in structure from these expectations.

The second issue is that there is a rather sharp distinction in \eqref{DPSmaster} between perturbatively
and nonperturbatively generated parton pairs, with the 2pGPD for the latter having a natural width in 
$\vect{r}$ space of order $\Lambda$ (as was discussed in section \ref{sec:2v1calc}). Does there exist 
some scale at which we can (approximately) regard all parton pairs in the proton as being `nonperturbatively 
generated' in this sense (as is assumed in \eqref{DPSmaster})? If so, what is the appropriate value for the 
scale (presumably it should be rather close to $\Lambda_{QCD}$)?

These issues are related in an essential way to the fact that we have cut the contribution from `1v1' 
graphs out of the DPS cross section entirely. It may therefore not be correct to entirely remove 
these graphs from the DPS cross section in this way. On the other hand, at present we do not have
a suitable alternative prescription for handling these graphs, and leave finding the appropriate
way of including the 1v1 graphs to future work.

\section{Conclusions} \label{sec:DPSXsecconc}

In this paper we have closely examined the contribution to the LO p-p DPS cross section from graphs 
in which two `nonperturbatively generated' ladders interact with two ladders that have been generated 
via a perturbative $1 \to 2$ branching process -- `2v1' graphs. We have presented a detailed 
calculation demonstrating that 2v1 graphs in which the two nonperturbatively generated ladders do
not interact with one another contribute to the LO p-p DPS cross section in the way originally written 
down by Ryskin and Snigirev \cite{Ryskin:2011kk}, and then later by Blok et al. \cite{Blok:2011bu} 
and Manohar and Waalewijn \cite{Manohar:2012jr, Manohar:2012pe}. We have also shown that 2v1 graphs
in which the `nonperturbatively generated' ladders exchange partons with one another contribute to 
the LO p-p DPS cross section, provided that this `crosstalk' occurs at a lower scale than the $1 \to 2$ 
branching on the other side of the graph. We have proposed a formula for the contribution from 2v1 graphs to
the LO DPS cross section, equation \eqref{2v1Xtalk}.

Crosstalk interactions between the two nonperturbatively generated ladders are suppressed by colour
effects -- for example, the `colour recombination' of figure \ref{fig:colourrecomb} is suppressed by a factor $1/(N_C^2 - 1)$.
This fact on it's own does not necessarily mean that the effect of crosstalk interactions in the 
2v1 diagrams is negligible. It was discovered in \cite{Bartels:1993ke} that crosstalk interactions could
lead to a sizeable increase in the cross section, even for rather small evolution lengths $\ln(Q^2/Q_0^2) 
\simeq 3$ and for not too small `final' $x$ values for the crosstalk $\simeq 10^{-3}$. However, we
pointed out that in the 2v1 diagrams, the typical $x$ values at which the $1 \to 2$ splitting occurs,
and the crosstalk finishes, are very much larger than those at the hard scale, and the $\mu^2$ value
of the splitting is much smaller than $Q^2$. For $Q = 10$ GeV, $x$ values at the hard scale $\gtrsim
10^{-3}$ correspond to $x$ values at the splitting $\gtrsim 0.1$, which is the region where PDFs
either do not change, or decrease with scale. In this region of $x$ the effect of crosstalk will either
be small, or will be to decrease the cross section. We obtain very similar conclusions when moving
to larger values of $Q^2$. Thus, except at 
exceedingly small $x$ values, we expect the effects of crosstalk interactions to be a reduction
of the 2v1 DPS cross section, which should be rather small at moderate $x$ values $\simeq 10^{-3}
- 10^{-2}$ corresponding to $x$ values at the splitting $\simeq 0.1$. A more precise statement 
than this requires a detailed numerical simulation of the 2v1 graphs, which is beyond the scope of
this paper.

We combined our formula for the 2v1 contribution to the DPS cross section \eqref{2v1Xtalk} with 
the suggestion that we made in \cite{Gaunt:2011xd} that 1v1 graphs should be completely removed 
from the DPS cross section to suggest a formula for the DPS cross section, equation \eqref{DPSmaster}.
Two potentially concerning features were identified in this equation, and the existence of these 
might indicate that completely removing the 1v1 graphs from the DPS cross section is not quite
the correct prescription. The determination of the appropriate manner of treating the 1v1 graphs
is left to future work.

\section*{Note Added}

After this paper was completed we learned of the published version of `What is Double 
Parton Scattering?' by Manohar and Waalewijn \cite{Manohar:2012pub} in which were corrected 
the errors of the original arXiv version \cite{Manohar:2012pe} that we discussed at the end of
section \ref{sec:crosstalk}. The discussion in that paper now appears to be in alignment 
with our own findings.

\section*{Acknowledgements}

We gratefully acknowledge discussions with J.~Stirling, and financial support from 
a Trinity College Senior Rouse Ball Studentship. All the figures in this paper were generated 
using \vb{JaxoDraw} \cite{Binosi200476}.


\providecommand{\href}[2]{#2}\begingroup\raggedright\endgroup

\end{document}